\documentclass[%
 preprintnumbers, 
 reprint,
 prd,
 amsmath,amssymb,
 aps,nofootinbib    
]{revtex4-2}
\usepackage[utf8]{inputenc}
\usepackage{xcolor}
\usepackage{graphicx}
\usepackage{dcolumn}
\usepackage{bm}
\usepackage{mathtools}
\usepackage{xcolor}
\usepackage{natbib}
\usepackage{graphicx}
\usepackage{amsmath}
\usepackage{hyperref}

\newcommand{\equ}[1]{\begin{eqnarray}#1\end{eqnarray}}

\newcommand{\tianshu}[1]{#1}

\newcommand{\apjl}{Astrophysical Journal, Letters}
\newcommand{\mnras}{Monthly Notices of the RAS}
\newcommand{\apjs}{Astrophysical Journal, Supplement}

\newcommand{\jcap}{Journal of Cosmology and    Astroparticle Physics}

\usepackage{comment}

\begin{document}
\preprint{LA-UR-25-25972}
\title[CFI in CCSNe]{The Effect of the Collisional Flavor Instability on Core-Collapse Supernova Models}
    
\author{Tianshu Wang}
\email{tianshuw@berkeley.edu}
\affiliation{Department of Physics, University of California, Berkeley, CA, 94720-7300 USA}
\author{Hiroki Nagakura}
\affiliation{Division of Science, National Astronomical Observatory of Japan, 2-21-1 Osawa, Mitaka, Tokyo 181-8588, Japan}
\author{Lucas Johns}
\affiliation{Theoretical Division, Los Alamos National Laboratory, Los Alamos, NM 87545, USA}
\author{Adam Burrows}
\affiliation{Department of Astrophysical Sciences, Princeton University, Princeton, NJ 08544}

\date{\today}

\begin{abstract}
We explore the effects of the neutrino collisional flavor instability (CFI) based on 1D and 2D core-collapse supernova (CCSN) simulations done using the sophisticated radiation-hydrodynamic code F{\sc ornax}. We compare the growth rates of homogeneous CFI (hCFI) modes calculated by numerically solving the multi-group dispersion relation to those calculated using the monochromatic approximation. We find that the widely-used monochromatic approximation leads to incorrect growth rates when applied in multi-group scenarios. As opposed to the $\sim10^5$ s$^{-1}$ values given by the monochromatic approximation, the actual growth rates of non-resonance multi-group hCFI are at most $\sim$200 s$^{-1}$ in all our models and they are too slow to affect CCSN outcomes. We adopt a BGK flavor conversion scheme in the simulations to include the effects of resonance-like hCFI. We find that the CCSN dynamics and neutrino emission properties are only weakly influenced, and the intrinsic stochasticity due to convection and neutrino-driven turbulence can naturally lead to comparable effects. Hence, our analysis of the non-resonance and resonance-like hCFI into CCSN simulations suggests that the effects of neutrino flavor conversion triggered by hCFI modes are in general small.
\end{abstract}

\begin{keywords}
\ supernovae - neutrino - neutrino oscillation 
\end{keywords}

\maketitle
\section{Introduction}
\label{sec:intro}
Neutrinos play an important role in the explosions of core-collapse supernovae (CCSNe). Supported by a number of sophisticated numerical simulations \citep{lentz:15,burrows_2019,vartanyan2019,muller_lowmass,stockinger2020,burrows_2020,bollig2021,sandoval2021,nakamura2022,vartanyan2023,burrows_correlations_2024}, the neutrino mechanism is widely accepted as the dominant explosion mechanism of the majority of CCSNe \citep{janka2012,burrows2013,burrows_vartanyan_nature,janka2025}. However, due to the high cost of such complex simulations, all these studies necessarily introduce approximations to some physical inputs and numerical algorithms. One of the greatest uncertainties in any modeling of CCSNe is the effect of neutrino flavor conversion, which is typically neglected in current simulations. Since the number densities and spectra of different neutrino flavors vary significantly, it is not impossible that neutrino flavor conversion could fundamentally alter the dynamics of explosion. Some attempts have been made to include flavor conversion through a phenomenological method \citep{ehring2023a,ehring_help_or_hinder2023,mori2025_ffc}, in which neutrino flavor equipartition is assumed once the matter density falls below a prescribed threshold -- treated as a free parameter. Depending upon the choice of the free parameter, the resulting explosion energy, if the model explodes, can vary by large factors \citep{mori2025_ffc}. Such large uncertainties highlight the need for further studies on the impact of neutrino flavor conversion in CCSNe.

To substantially influence CCSN dynamics, neutrino flavor conversion must occur in the post-shock region before the onset of the explosion, which spans only $\sim100 - 150$ kilometers (km). 
Therefore, flavor instabilities with growth rates higher than $\sim10^3$ s$^{-1}$ are needed.
In the core region of CCSNe where neutrino number densities are high, neutrino self-interactions can induce collective flavor oscillations on very short timescales \citep{Pantaleone1992,duan2010}. Several types of such collective flavor conversions have been proposed. One is known as fast flavor conversion (FFC) which arises when the angular distribution of neutrino-flavor-lepton-number (NFLN) crosses zero \citep{morinaga2022,dasgupta2022}. The presence of such NFLN crossings has been confirmed in a number of CCSN simulations \citep{abbar2019,Nagakura2019_ffc,Delfan2020,abbar2020,glas_osc_2020,abbar2020b,morinaga2020,capozzi2021,abbar2021,johns_nagakura2021,nagakura_johns2021,richers2021,richers2021b,harada2022,akaho2023,Liu2023b,akaho2024}. Although the FFC can evolve on nanosecond or shorter timescales \citep{volpe2015,richers2019,nagakura2023,nagakura2023b,volpe2024,xiong2024} and significantly alter the properties of emitted neutrinos, recent self-consistent CCSN simulations incorporating an approximate FFC solver suggest that FFC has only a minimal impact on the overall CCSN dynamics \citep{wang2025}. This is because the dominant region of FFC tends to lie at relatively large radii (mostly in the pre-shock region), where neutrinos are already decoupled from matter. At smaller radii, where flavor conversion could influence the hydrodynamics, the neutrino angular distributions are nearly isotropic and angular crossings hardly occur.

Another flavor conversion mode has been pointed out by \citep{johns_collisional2023}, known as the collisional flavor instability ({CFI}). This instability can be triggered for isotropic neutrino distributions and, thus, it can occur at smaller radii compared to the FFC. Starting from the quantum kinetic equation (QKE) of neutrinos, the dispersion relation governing this instability can be derived by treating neutrino coherence as a perturbation \citep{Padilla2022,Lin2023,Xiong2023,Liu2023,akaho2024,Froustey2025}. Many works focus on homogeneous (vanishing wave vector) CFI {(hereafter, hCFI)} modes in {isotropic neutrino backgrounds}. Under the assumption of monochromatic neutrinos, the growth rate can be solved analytically, and a few studies on continuous neutrino spectra claim that the monochromatic formulae provide a good approximation to the multi-group results \citep{Lin2023,Xiong2023,Liu2023}. Consequently, previous analysis of hCFI in CCSNe and binary neutron star (BNS) mergers typically reduce the multi-group neutrino spectra to integrated quantities and substitute them into the monochromatic formulae \citep{Liu2023b,akaho2024,nagakura2025}. Using this approach, the maximum hCFI growth rates in CCSNe have been estimated to be on the order of $\sim 10^5$ s$^{-1}$, which is comparable to the neutrino escape time through the hCFI unstable region located roughly between $10^{10}$ and $10^{12}$ g cm$^{-3} $\citep{Liu2023b,akaho2024}. Based on such monochromatic estimations, it seemed possible for hCFI to substantially change CCSN models.

However, as we will demonstrate in this paper, using the monochromatic formulae in multi-group scenarios can be problematic. We find that the analytical derivations presented in \citep{Lin2023,Xiong2023} are incomplete and can fail under certain conditions. The numerical tests of the monochromatic approximation in multi-group contexts \citep{Liu2023} do not explore the relevant regions of parameter space and, therefore, cannot be regarded as comprehensive validation. In this work, we identify the issues in the analytical justifications of \citep{Lin2023,Xiong2023} and present a numerical example in which the monochromatic formulae produce qualitatively incorrect results. We then revisit the identification of hCFI regions in CCSN simulations done using our radiation-hydrodynamic code F{\sc ornax}, and we find that the monochromatic formulae largely overestimate the impact of non-resonance hCFI {in our models}. In fact, the size of non-resonance hCFI unstable regions is overestimated by a large factor, and the maximum non-resonance hCFI growth rate is overestimated by almost three orders of magnitude (when compared with more accurate results obtained by numerically solving the full multi-group dispersion relation). The actual non-resonance hCFI growth rates don't exceed $\sim200$ s$^{-1}$ and the associated growth timescale is significantly longer than the neutrino advection time through the unstable region. 
Our result suggests that the effects of non-resonance hCFI in CCSNe are negligible.

The growth rates of the resonance-like hCFI, on the other hand, are still well-approximated by the monochromatic formulae. We confirm the findings in \citep{akaho2024} that the resonance-like hCFI occur in high matter density regions. The resonance condition requires $A\equiv \sqrt{2}G_F(n_{\nu_e}-n_{\nu_x}-n_{\bar{\nu}_e}+n_{\bar{\nu}_x})=0$, which limits the resonance-like hCFI to very thin layers in the proto-neutron star (PNS). Despite the small size of this unstable region, the resonance-like hCFI growth rate {can be several orders of magnitudes higher than the non-resonance case} and, thus, its effects cannot simply be ignored. We use the recently implemented BGK scheme in F{\sc ornax} \citep{wang2025} to incorporate resonance-like hCFI self-consistently in CCSN simulations. We find that resonance-like hCFI only introduce minor effects on CCSN dynamics and the neutrino properties measured at large radii, which is comparable to the effects of the intrinsic stochasticity due to convection and neutrino-driven turbulence. The remaining uncertainties in the nuclear equation of state \citep{couch2013,yasin2020,Boccioli2022} and in the structures of the initial model progenitors \citep{swbj16,sukhbold2018} likely swamp this effect. Hence, our analysis of the non-resonance and resonance-like hCFI into CCSN simulations suggests that the effects of neutrino flavor conversion triggered by hCFI modes are in general small. 

This paper is arranged as follows. In Section \ref{sec:method} we describe details of our computational approach. Then in Section \ref{sec:hCFI_gr} we provide the monochromatic formulae for the hCFI growth rates and discuss their failure in the more realistic multi-group context. Our analysis of the presence and effects of hCFI in F{\sc ornax} CCSN models is provided in Section \ref{sec:result}. Finally, we summarize our findings and discuss the limitations in Section \ref{sec:conclusion}.

\section{Method}
\label{sec:method}
The sophisticated code F{\sc ornax}, described in detail in \citep{skinner2019,burrows_40,vartanyan2019}, by default uses a 3-species neutrino scheme which combines all heavy neutrino species ($\nu_\mu$, $\nu_\tau$, $\bar{\nu}_\mu$, $\bar{\nu}_\tau$) into one single neutrino type in the code. In this study, however, we implement in F{\sc ornax} a 4-species scheme: $\nu_\mu$ and $\nu_\tau$ are merged into $\nu_x$, while $\bar{\nu}_\mu$ and $\bar{\nu}_\tau$ are merged into $\bar{\nu}_x$. This allows us to more accurately track the effects of neutrino flavor conversion and avoids the potential indirect mixing between neutrinos and anti-neutrinos \citep{wang2025,akaho2025}. We ignore muons and tauons in the simulation, and we assume that $\nu_x$ and $\bar{\nu}_x$ have the same emissivity/opacity. Therefore, the only source of $\nu_x$ and $\bar{\nu}_x$ difference is flavor mixing. By default, we use twelve neutrino energy groups logarithmically distributed between 1 and 300 MeV for each species. We have also done tests in 1D with 40 energy groups and see only very minor differences. With the 12 energy group data, we have also tried several different interpolation methods: (a) use the original 12 energy groups, (b) use linear interpolation to 100 points between 0 and 100 MeV, and (c) use cubic spline interpolation to 100 points between 0 and 100 MeV. The change in the hCFI growth rate calculation is on the level of a few percent, so our results are insensitive to the integration scheme.
The neutrino-matter microphysics used in the classical physics sector can be found in \citep{2006NuPhA.777..356B} and \citep{2020PhRvD.102b3017W}. \tianshu{These include the super–allowed charged–current absorptions of $\nu_e$ and $\bar{\nu}_e$ neutrinos on free nucleons; neutral–current scattering off of free nucleons, alpha particles, and nuclei; neutrino–electron scattering; neutrino–nucleus absorption (and its inverse); and the inverses of various neutrino production processes such as nucleon–nucleon bremsstrahlung and pair production. Many-body corrections to the axial-vector part of the neutrino-nucleon scattering rate are taken from \citep{Horowitz2017}. Weak magnetism and recoil corrections to scattering and absorption rates off nucleons \`a la \citep{horowitz2002} are employed. Energy redistribution in ``inelastic" scattering of neutrinos off electrons is addressed using the prescriptions described in \citep{Thompson2003}, while the corresponding energy redistribution in inelastic scattering off nucleons is handled using a Kompaneets method \citep{2020PhRvD.102b3017W}.}

We carry out 1D and 2D simulations of 9, 12.25, 14, 18, 20, and 25 $M_\odot$ ZAMS-mass models. These are fiducial reference models in which flavor conversion is turned off and are used to identify the hCFI unstable regions. The 9 and 12.25 $M_\odot$ progenitors are taken from \citep{swbj16}, while the 14, 18, 20, and 25 $M_\odot$ progenitors are taken from \citep{sukhbold2018}. All progenitors are solar metallicity. There are always 1024 radial zones, while the 2D simulations have $128$ zones along the $\theta$ direction. The outer boundary is set at 30,000 kilometers (km) and the inner radial zone is 0.5 km wide. The SFHo nuclear equation of state \citep{2013ApJ...774...17S} is employed.

Based on the analysis of these reference models, we select a few models and redo the simulation including the hCFI effects. The flavor conversion triggered by hCFI is handled using a BGK formalism \citep{Nagakura2024,wang2025}. Given the instability growth rate $\sigma$, the change in the neutrino distribution function per simulation timestep $\Delta t$ is
\equ{
f'_{\nu_\alpha}(E) - f_{\nu_\alpha}(E) = -(1-e^{-\sigma \Delta t})(f_{\nu_\alpha}(E)-f^a_{\nu_\alpha}(E))\, ,\nonumber\\
}
where $f^a_{\nu_\alpha}(E)$ is the asymptotic state of flavor conversion. As will be shown in Section \ref{sec:hCFI_gr}, only the resonance-like hCFI influences CCSN dynamics and, thus, we apply such a BGK scheme only to the resonance-like hCFI regions. Technically, the BGK scheme is applied to regions where the monochromatic hCFI growth rate $\sigma>10^7$ s$^{-1}$\footnote{\tianshu{This $\sigma>10^7$ s$^{-1}$ threshold is an empirical criterion for resonance-like hCFI. Due to the high computational cost, we can't afford the accurate multi-group growth rate calculation on-the-fly, so we use the monochromatic growth rate formula which only works in the resonance-like case. This is why such a resonance criterion is needed.}}. The typical resonance-like hCFI growth rates in our simulations are $\sim10^{9}$ s$^{-1}$. We choose to use the monochromatic formulae in the BGK scheme because of its simplicity. Although the monochromatic formulae fail in the non-resonance case, they are able to give the right results in the resonance-like case as shown in Section \ref{sec:hCFI_gr}. Given the high growth rates of the resonance-like hCFI, the flavor asymptotic states will be instantly achieved every hydrodynamic (Courant) timestep ($\sim10^{-6}$s). We note that \citet{akaho2025} finds that the BGK scheme tends to overestimate the effects of FFC with $\sim10^{-6}$s time steps. If the same is true for CFI, results in this paper can be regarded as an upper limit on the effects of CFI-triggered flavor conversion and doesn't change our conclusion.

Although analytical expressions for the asymptotic states in a monochromatic neutrino system are provided by \citet{Froustey2025}, they cannot be easily generalized to the multi-group case. 
Multi-group QKE simulations have shown that flavor equipartition and flavor swap could both be the asymptotic flavor state, depending upon the assumed form of the collisional term and initial conditions \citep{kato2024,Zaizen2025}.
We choose flavor equipartition to be the asymptotic states of flavor conversion:
\equ{
&&f^a_{\nu_e}(E)=f^a_{\nu_x}(E)=\frac{f_{\nu_e}(E)+2f_{\nu_x}(E)}{3}\nonumber\,,\\
&&f^a_{\bar{\nu}_e}(E)=f^a_{\bar{\nu}_x}(E)=\frac{f_{\bar{\nu}_e}(E)+2f_{\bar{\nu}_x}(E)}{3}\,.
}
This is because flavor equipartition is more numerically stable than flavor swap in resonance-like case. 
Although our results seem not to be sensitive to the choice of asymptotic states due to the overall weak effects of hCFI, it will be useful to repeat this type of simulation with better subgrid models.

\begin{figure*}[t]
    \centering
    \includegraphics[width=0.48\textwidth]{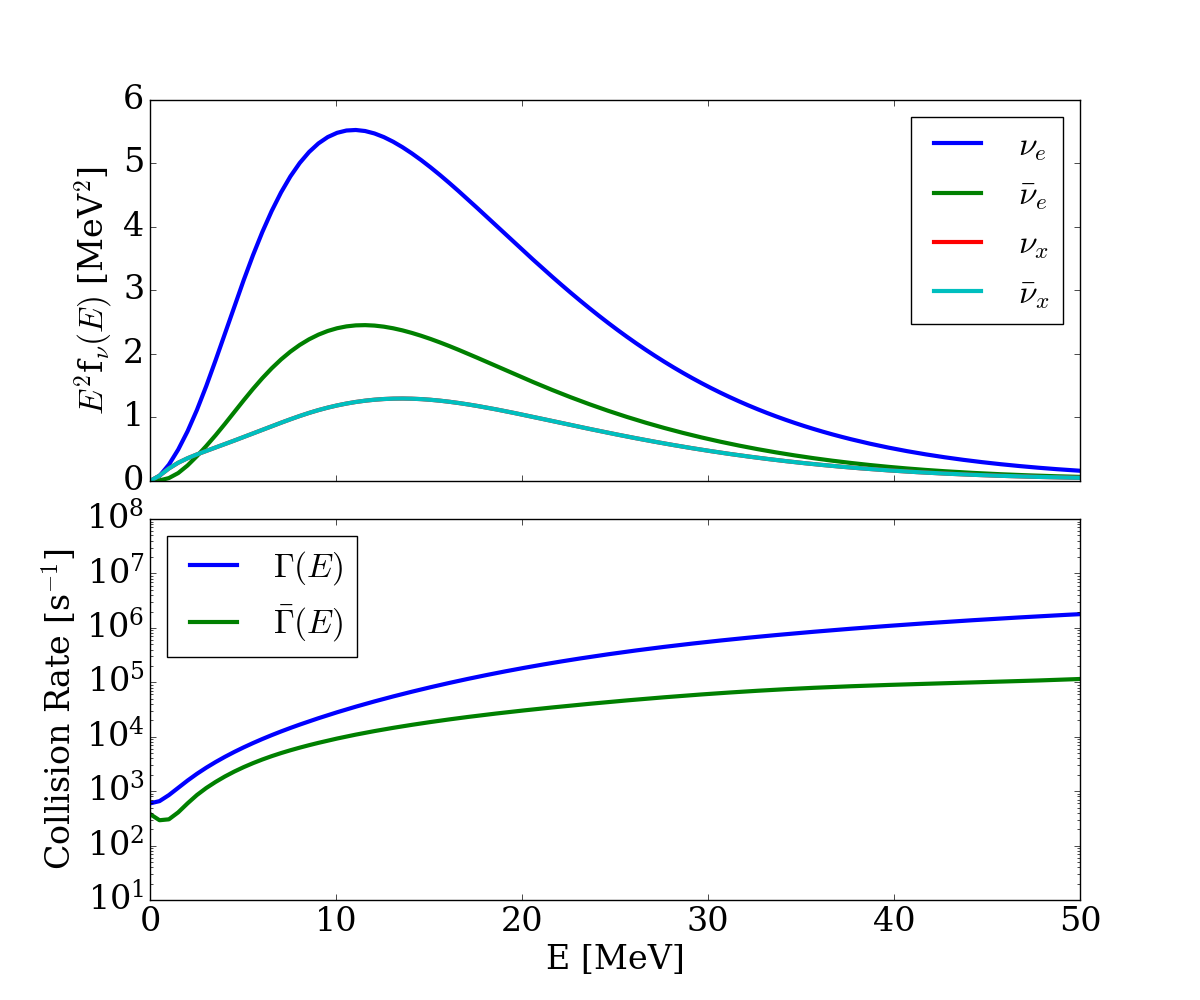}
    \includegraphics[width=0.48\textwidth]{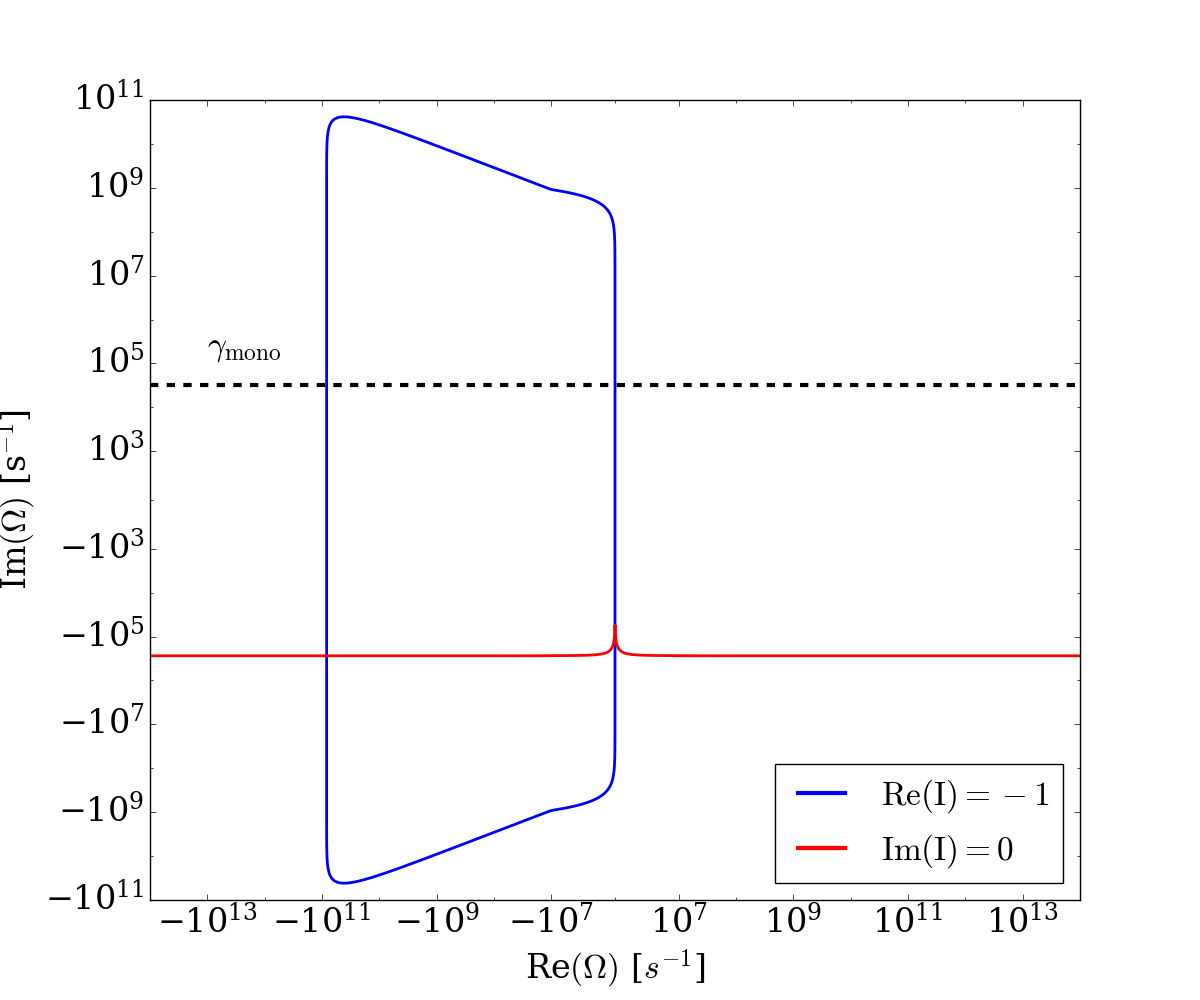}
    \caption{A fundamental failure of the monochromatic formulae applied to a multi-group case. {\bf Left:} normalized neutrino number spectra and the collision rates taken from the 18$M_\odot$ 1D simulation at a radius of 37km and 120 milliseconds post-bounce. The local density, temperature, and $Y_e$ are $1.016\times10^{12}$ g cm$^{-3}$, 7.45 MeV, and 0.1426, respectively. The $\nu_x$ and $\bar{\nu}_x$ distribution functions are identical. Based on the monochromatic formulae, this location shows the strongest hCFI growth rate at that time. {\bf Right:} Contour lines for Re(I)=$-$1 and Im(I)=0 in Eq. \ref{equ:dispersion_relation}. The cross points of the contours give the solutions to the dispersion relation. The horizontal dashed line depicts the maximum monochromatic growth rate. All numerical solutions have negative imaginary parts and there is no unstable mode. Although we show only the isotropy-preserving branch (I$=-1$), the same is true for the isotropy-breaking branch (I$=3$) because the growth rates are hard limited by the imaginary part for the equation (Im(I)=0). However, the monochromatic formulae give a growth rate of a few $10^4$ s$^{-1}$, which is qualitatively different from the correct solution.}
    \label{fig:example}
\end{figure*}

\section{CFI Growth Rate Calculation}
\label{sec:hCFI_gr}
The evolution of dense neutrino gases in CCSNe is described by the quantum kinetic equation (QKE) 
\equ{
(\partial_t+\vec{v}\cdot\nabla)\rho&=&-i[\mathcal{H}_{\nu\nu},\rho]+\mathcal{C}[\rho]\, ,
}
where $\rho$ denotes the neutrino density matrix, $\mathcal{C}$ the collision term, and $\mathcal{H}_{\nu\nu}=\sqrt{2}G_F\int_{-\infty}^{\infty}\frac{E'^2dE'}{2\pi^2}\int_{\Omega'_\nu}\frac{d\Omega'_\nu}{4\pi}v'_{\mu}v^{\mu}\rho'$ a neutrino self-interaction Hamiltonian. Here, natural units ($\hbar=c=1$) are adopted. We assume here that the vacuum and matter oscillations can be treated as sources of perturbations, and we employ the common convention $\rho(-E)=-\bar{\rho}(E)$. We further assume that the background neutrino gases are locally isotropic and homogeneous \citep{johns_collisional2023,Lin2023,Liu2023,Liu2023b,akaho2024}, and we focus on hCFI modes with vanishing wave vectors. Although this assumption may fail in the optically-thin regions where the neutrino angular distribution becomes forward-peaked, the longer neutrino-matter interaction timescale in those regions leads to negligible hCFI growth anyway \citep{akaho2024}. 

The collision term $\mathcal{C}[\rho]$ can be written in the relaxation approximation as $\mathcal{C}[\rho]=\frac{1}{2}\{{\rm diag(\Gamma_e, \Gamma_x),\rho_{\rm eq}-\rho}\}$, where $\Gamma_\alpha$ is the collision rate and $\rho_{\rm eq}$ is the equilibrium state \citep{johns_collisional2023}. To calculate $\Gamma_\alpha$, we include all emission/absorption interactions implemented in F{\sc ornax} \citep{2006NuPhA.777..356B}, while scattering processes are neglected when calculating the hCFI effects. \tianshu{This is because we assume isotropic neutrino distribution and scattering terms are exactly canceled \citep{Liu2023b,akaho2024}. }

By treating the off-diagonal terms of the density matrix as perturbations, the mean-field density matrix of neutrino flavor can be written as \citep{Liu2023,akaho2024}
\equ{
\rho&\approx&\frac{f_{\nu_e}+f_{\nu_x}}{2}+\frac{f_{\nu_e}-f_{\nu_x}}{2}
\begin{pmatrix}
1 & S \\
S^* & \tianshu{-1} 
\end{pmatrix}\, ,
}
where $|S|\ll1$ is the neutrino flavor coherence. Similar to the density matrix convention, we have $f_{\nu_\alpha}(-E)=-\bar{f}_{\nu_\alpha}(E)$. After linearizing the QKE, one can derive the dispersion relation for the homogeneous mode \citep{Lin2023,Liu2023}:
\equ{
{\rm I}&\equiv&\sqrt{2}G_F\int_{-\infty}^{\infty} \frac{E^2dE}{2\pi^2}\frac{f_{\nu_e}(E)-f_{\nu_x}(E)}{\Omega+i\Gamma(E)}=-1\,\, {\rm or}\,\, 3 \, , \label{equ:dispersion_relation}
}
where $\Gamma(E)=(\Gamma_e(E)+\Gamma_x(E))/2$ and $\Gamma_\alpha(-E)=\bar{\Gamma}_\alpha(E)$.
The imaginary part of the frequency, ${\rm Im}(\Omega)$, is the growth rate of hCFI modes. The solution branches ${\rm I}=-1$ and ${\rm I}=3$ are called the isotropy-preserving branch and isotropy-breaking branch, respectively \footnote{\tianshu{Although Eq. \ref{equ:dispersion_relation} provides a good way to visualize the roots (as done in Figures \ref{fig:example} and \ref{fig:failure}), it is not the most robust numerical method to determine all roots. Common root finder algorithms will miss some solutions unless good initial guesses are provided. Therefore, the growth rates are calculated by solving the eigenvalues of the linearized QKE equation $(\Omega+i\Gamma(E))Q(\Omega,E,\hat{n})=-\sqrt{2}G_F(f_{\nu_e}(E)-f_{\nu_x}(E))\int^{+\infty}_{-\infty}\frac{E'^2dE'}{(2\pi)^3}\int d\hat{n}' (1-\hat{n}\cdot\hat{n}')Q(\Omega,E',\hat{n}')$. The two methods are mathematically equivalent and the results agree with each other very well.}}. 

\subsection{Monochromatic Growth Rates}
If the neutrino distributions are monochromatic,
\equ{f_{\nu_e}-f_{\nu_x}&=&\frac{2\pi^2}{\sqrt{2}G_F E^2}[g\delta(E-\varepsilon)-\bar{g}\delta(E-\bar{\varepsilon})]\, ,}
where $g$, $\bar{g}$, $\varepsilon$, and $\bar{\varepsilon}$ are parameters and the dispersion relation Eq. \ref{equ:dispersion_relation} can be simplified to
\equ{
&&\frac{g}{\Omega+i\Gamma}-\frac{\bar{g}}{\Omega+i\bar{\Gamma}} = -1\,\, {\rm or}\,\, 3\, .
}
The solution to this equation is
\equ{
\Omega_\pm&=&-A-i\gamma\pm\sqrt{A^2-\alpha^2+i2G\alpha}
}
for the isotropy-preserving (I=$-$1) branch, and
\equ{
\Omega_\pm&=&-A/3-i\gamma\pm\sqrt{(A/3)^2-\alpha^2-i2G\alpha/3}
}
for the isotropy-breaking (I=$3$) branch, where
\equ{
G = \frac{g+\bar{g}}{2}, \,\, A = \frac{g-\bar{g}}{2}, \,\, \gamma=\frac{\Gamma+\bar{\Gamma}}{2}, \,\, \alpha=\frac{\Gamma-\bar{\Gamma}}{2}\, .
}
The $\Omega_+$ and $\Omega_-$ solutions are referred to as the plus and minus modes. Since the isotropy-preserving mode has higher growth rates \citep{Liu2023,Zaizen2025}, we will ignore the isotropy-breaking modes hereafter\footnote{Although this growth rate comparison is done under the monochromatic assumption, we will see later in the multi-group case that the maximum growth rates are more strictly constrained by the imaginary part of the dispersion relation, so the differences between isotropy-preserving and isotropy-breaking modes don't change our conclusions.}. The asymptotic behavior is:
\equ{
\Omega_\pm &\approx 
\begin{cases} 
      -A-i\gamma\pm(|A|+iG\alpha/|A|)  & A^2\gg |G\alpha| \\
      -A-i\gamma\pm\sqrt{i2G\alpha} & A^2 \ll G|\alpha| \\
\end{cases}\,.\nonumber\\
\label{equ:mono_growth}
}

{Another form of the monochromatic formulae can be found in \citet{Lin2023}. We omit the derivation but list the result here:
\equ{
{\rm Im}(\Omega_+)&=&\frac{\Gamma \bar{g}-\bar{\Gamma}g}{g-\bar{g}}\nonumber\\
{\rm Im}(\Omega_-)&=&-\frac{\Gamma g-\bar{\Gamma}\bar{g}}{g-\bar{g}}\,,
\label{equ:mono_growth_1}
}
which is equivalent to the $A^2\gg G|\alpha|$ (non-resonance) case of Eq. \ref{equ:mono_growth}.
}

\tianshu{In CCSNe, the neutrino numbers and collision rates follow $n_{\nu_e}>n_{\bar{\nu}_e}>n_{\nu_x}\approx n_{\bar{\nu}_x}$ and $\Gamma_{e}>\bar{\Gamma}_{e}>\Gamma_{x}\approx \bar{\Gamma}_{x}$, while the high neutrino number densities lead to $\sqrt{2}G_Fn_\nu\gg\Gamma$. Therefore, CCSNe always have $G>0$ and $\alpha>0$, and in non-resonance cases ($A\neq0$) $A^2\gg G|\alpha|$ also holds.}

\begin{figure*}
    \centering
    \includegraphics[width=0.48\textwidth]{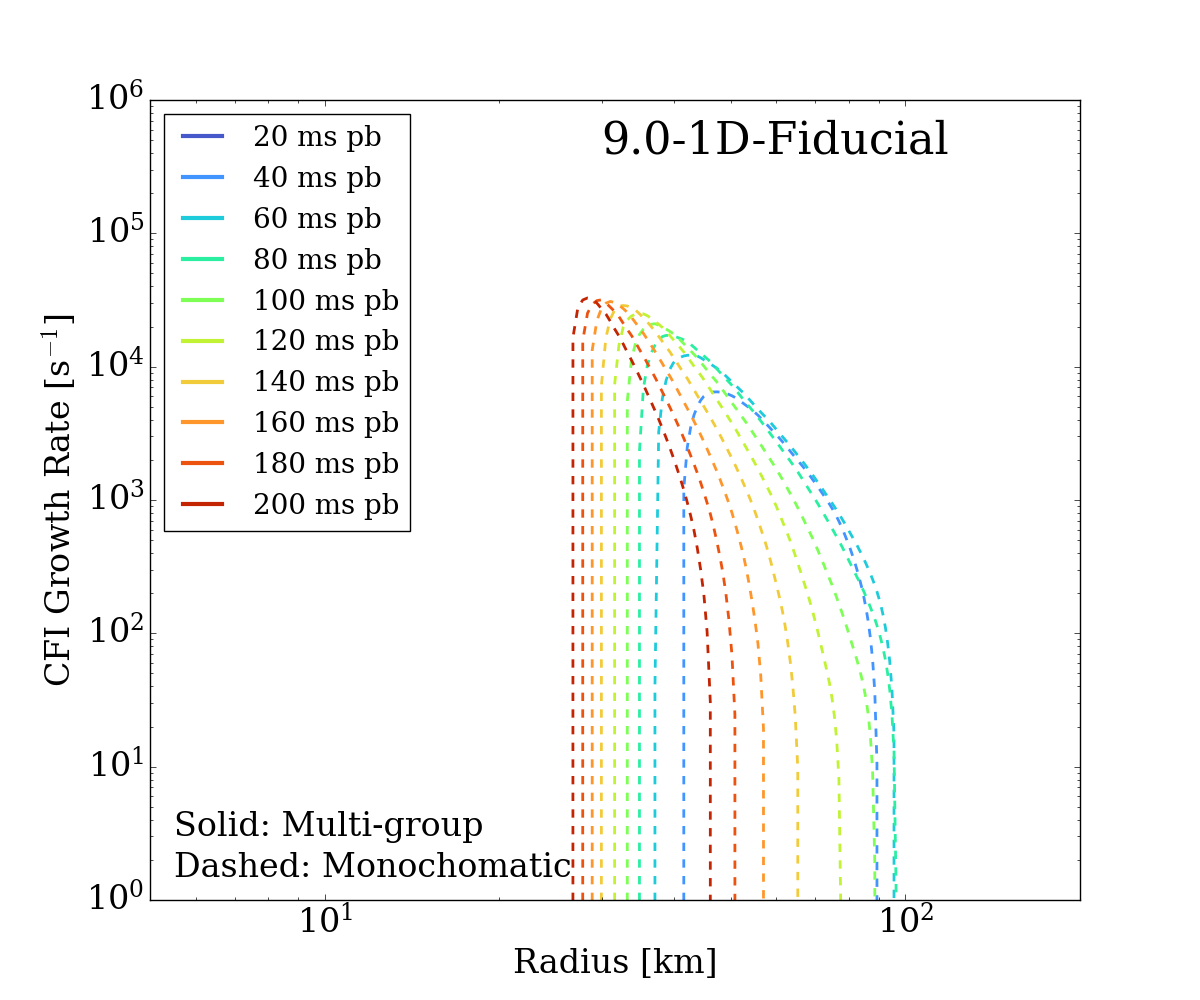}
    \includegraphics[width=0.48\textwidth]{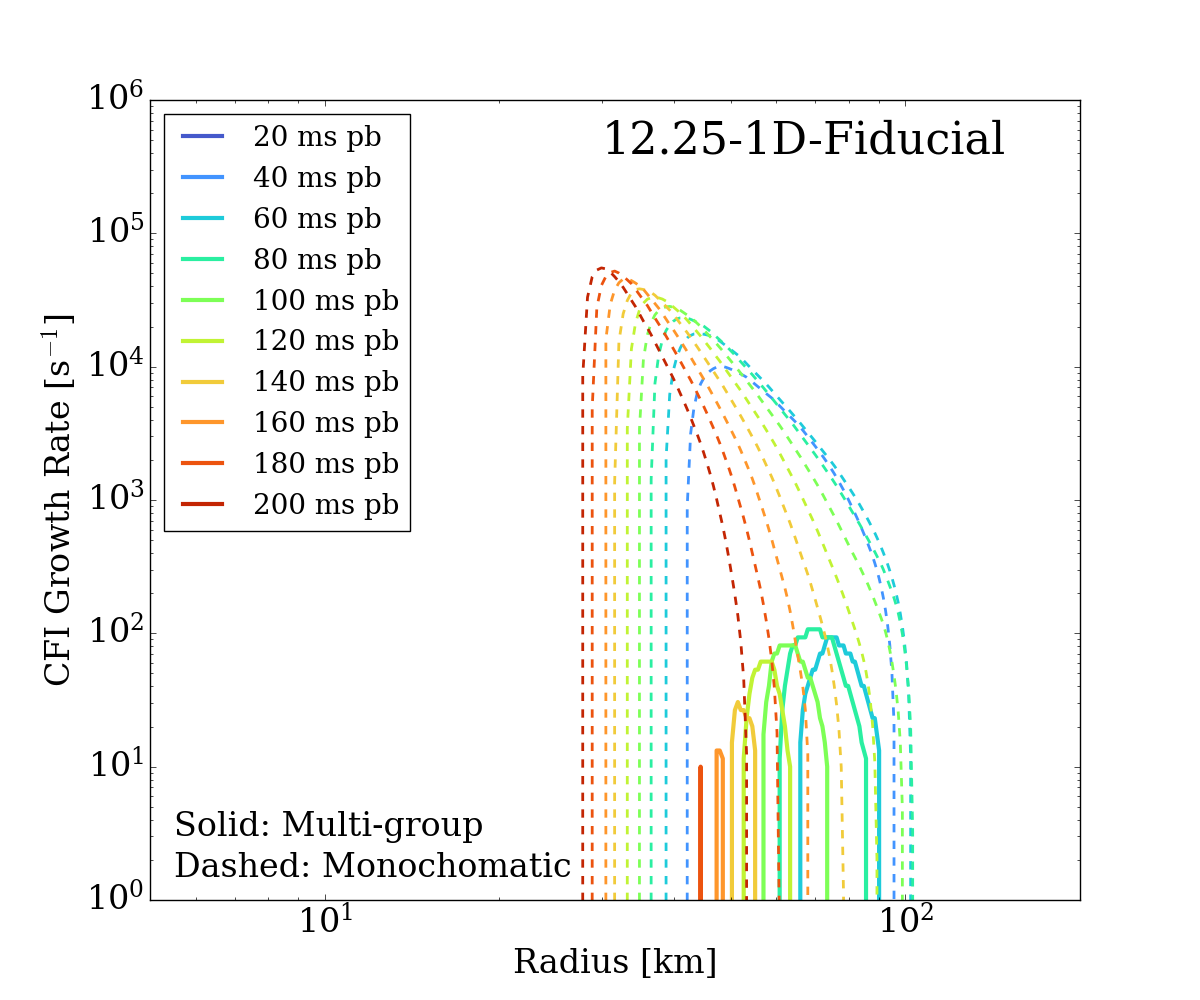}
    \includegraphics[width=0.48\textwidth]{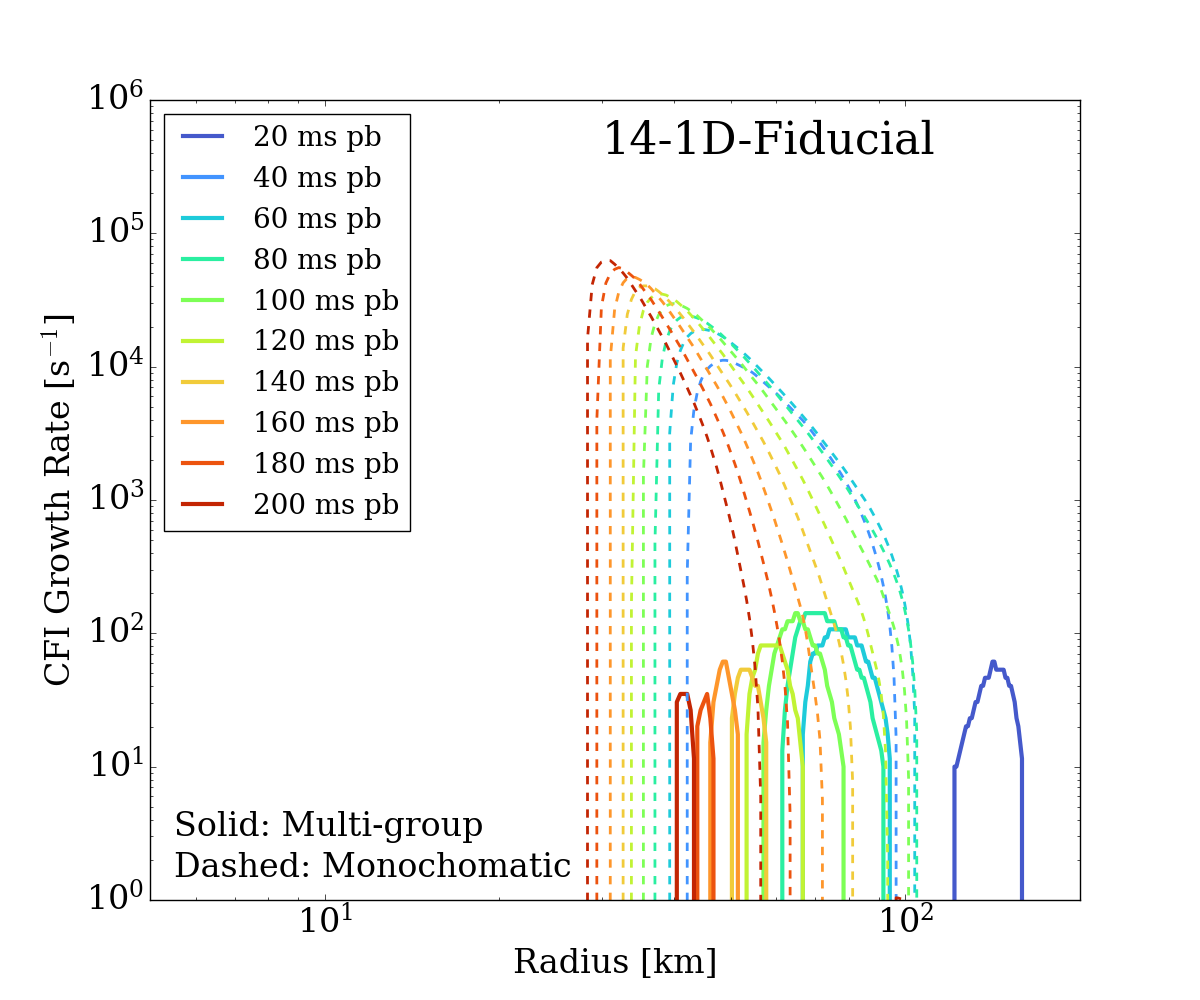}
    \includegraphics[width=0.48\textwidth]{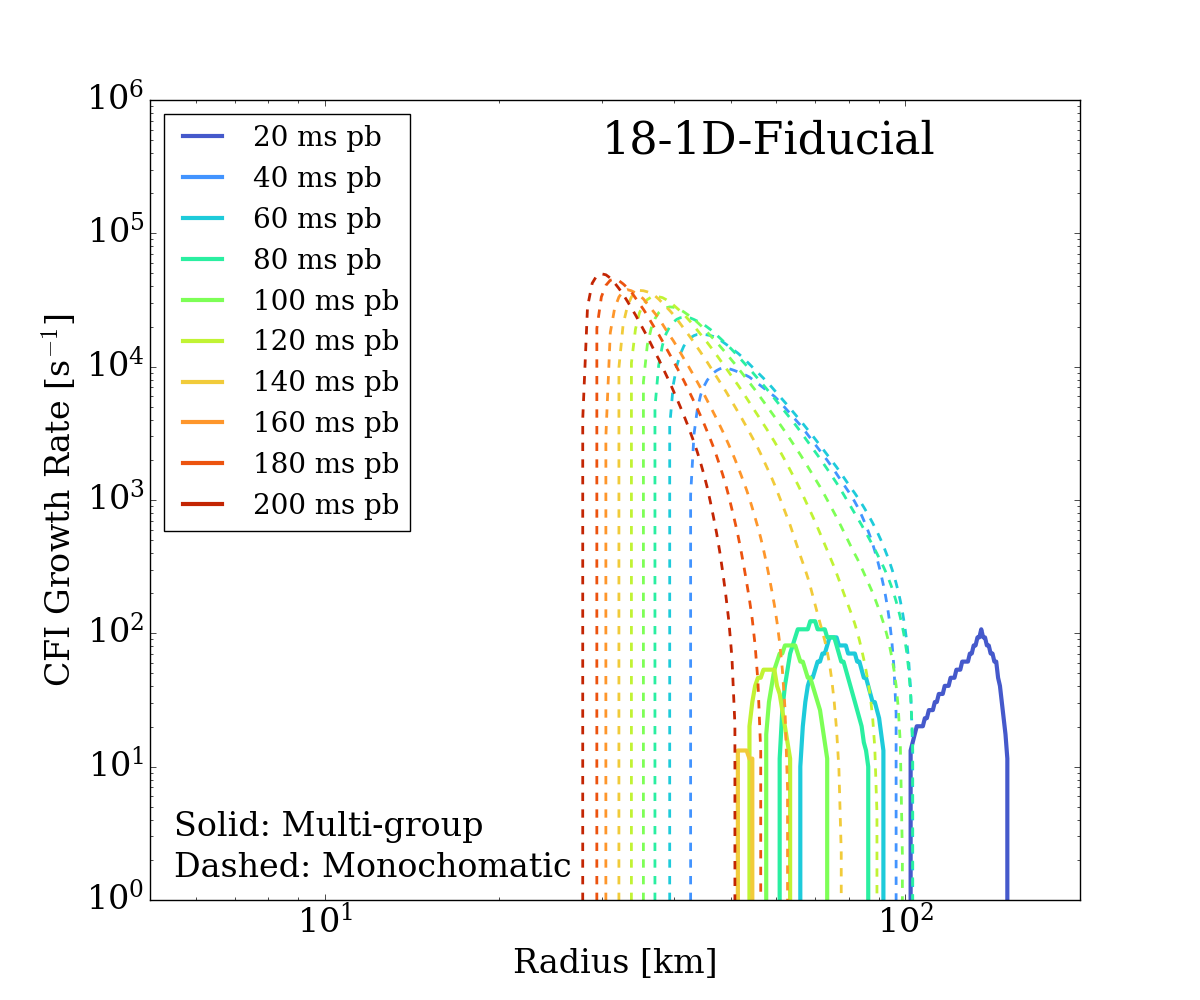}
    \includegraphics[width=0.48\textwidth]{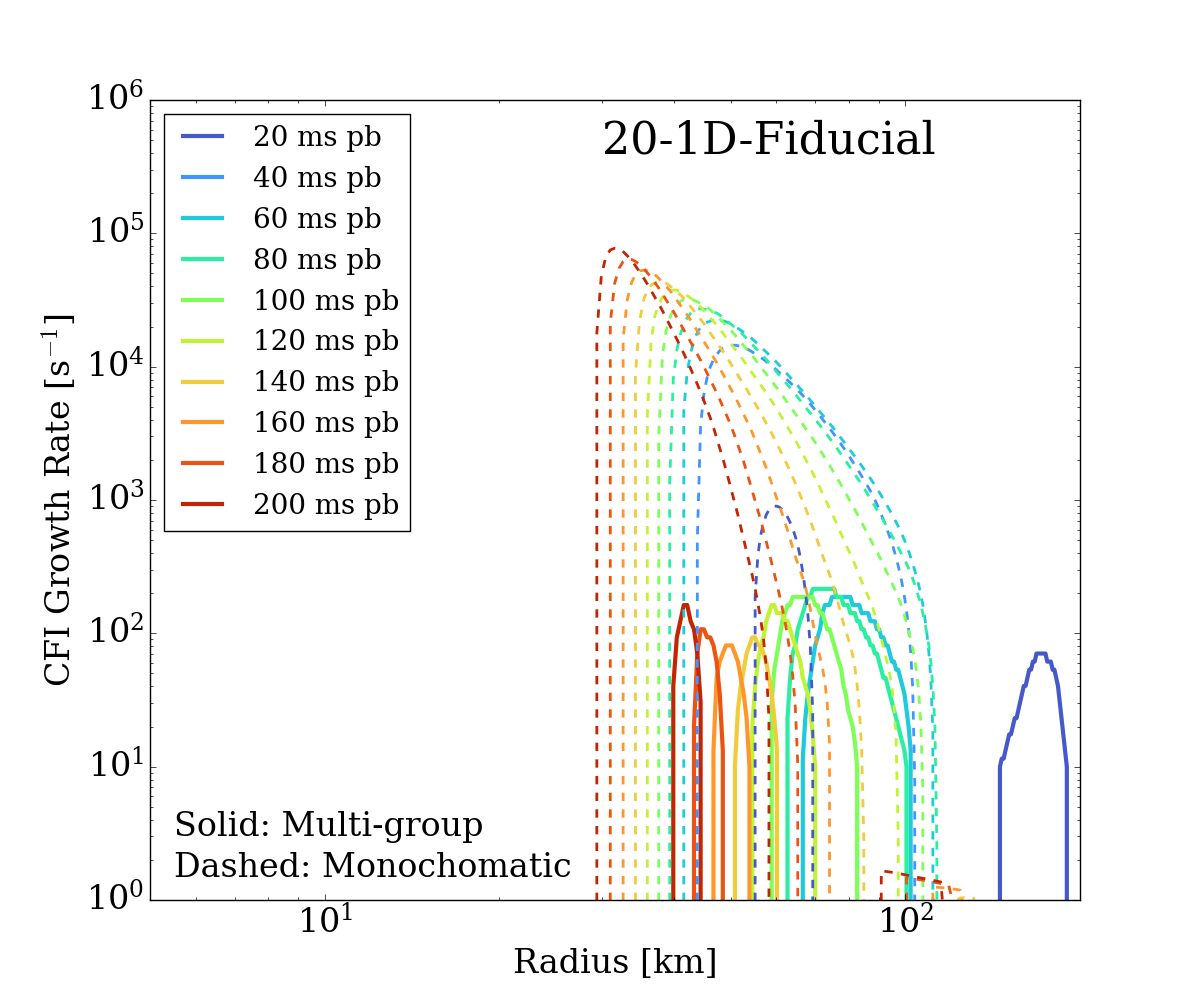}
    \includegraphics[width=0.48\textwidth]{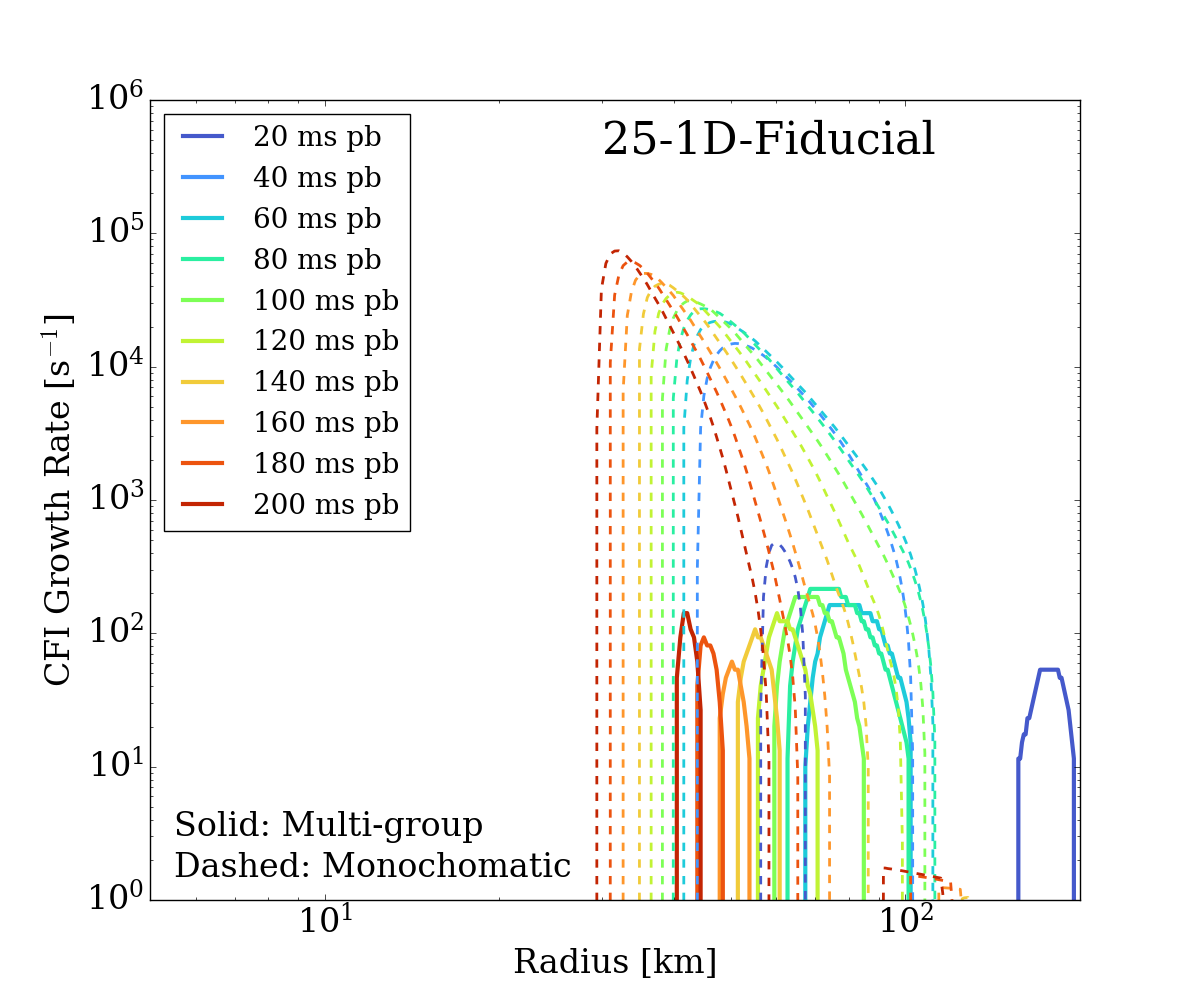}
    \caption{hCFI growth rate profiles at several snapshots of the 9, 12.25, 14, 18, 20, and 25 $M_\odot$ 1D fiducial models in the first 200 milliseconds post-bounce. Solid curves depict the growth rates calculated by numerically solving the multi-group dispersion relation, while dashed curves represent the monochromatic growth rates. All the hCFI regions shown are non-resonant. We don't see any resonance-like behavior in 1D models. In all our models, the monochromatic formulae overestimate the maximum hCFI growth rates by almost three orders of magnitudes. With the more accurate multi-group method, none of the models shows non-resonant hCFI growth rates above $\sim$200 s$^{-1}$. The actual hCFI growth rates in the 9 $M_\odot$ model are even lower than $1$ s$^{-1}$, below the lower-limit of the y-axis of the plot. Since the size of hCFI regions is at most a few tens of kilometers, the neutrino flux timescale through the unstable region ($\sim$10$^{-4}$ s) is much shorter than the timescale of hCFI growth. As a result, the effect of flavor conversion triggered by non-resonant hCFI can be neglected.}
    \label{fig:growth_rates}
\end{figure*}
\subsection{An Assessment of the Monochromatic Treatment in Multi-Group Cases}
\label{sec:criticism}
Although the above formulae are derived under the monochromatic assumption, it was thought that they would also work in multi-group cases \citep{Lin2023,Xiong2023,Liu2023,akaho2024}. The parameters $g$ and $\bar{g}$ are connected to the continuous neutrino distribution functions via 
\equ{
g&=&\sqrt{2}G_F(n_{\nu_e}-n_{\nu_x})\nonumber \\
\bar{g}&=&\sqrt{2}G_F(n_{\bar{\nu}_e}-n_{\bar{\nu}_x})\, ,
} 
where $n_{\nu_\alpha}=\int_0^\infty\frac{E^2dE}{2\pi^2}f_{\nu_\alpha}(E)$ is the neutrino number density. There are two slightly different averaging schemes for collision rates in the literature \citep{Liu2023,akaho2024}, {and we refer to them as methods A and B}. 
{Method A} \citep{Liu2023,Lin2023,Xiong2023} is to calculate $\Gamma$ and $\bar{\Gamma}$ using 
\equ{
\Gamma&=&\frac{\int_0^\infty E^2dE [f_{\nu_e}(E)-f_{\nu_x}(E)]\Gamma(E)}{\int_0^\infty E^2dE (f_{\nu_e}(E)-f_{\nu_x}(E))}\nonumber\\
\bar{\Gamma}&=&\frac{\int_0^\infty E^2dE [f_{\bar{\nu}_e}(E)-f_{\bar{\nu}_x}(E)]\bar{\Gamma}(E)}{\int_0^\infty E^2dE (f_{\bar{\nu}_e}(E)-f_{\bar{\nu}_x}(E))}\, ,
} 
where $\Gamma(E)=(\Gamma_e(E)+\Gamma_x(E))/2$, {and substitute these integrated quantities into Eq. \ref{equ:mono_growth_1}.} 

Method B \citep{akaho2024} is to calculate the average collision rates separately for each neutrino species, 
\equ{
\Gamma_\alpha=\frac{\int_0^\infty E^2dE f_{\nu_\alpha}(E)\Gamma_\alpha(E)}{\int_0^\infty E^2dE f_{\nu_\alpha}(E)}
}
and express $\Gamma$ and $\bar{\Gamma}$ as $\Gamma=(\Gamma_e+\Gamma_x)/2$ and $\bar{\Gamma}=(\bar{\Gamma}_e+\bar{\Gamma}_x)/2$, {and substitute these integrated quantities into Eq. \ref{equ:mono_growth}.} 
In our tests, the two averaging schemes qualitatively agree with each other, {but for a clearer view we discuss them separately below.}

The justifications for {Method A} have been provided both theoretically and numerically in \citet{Xiong2023,Lin2023}. The justifications for {Method B} have been provided only numerically in \citet{Liu2023}.
However, as we will show next, {the theoretical justifications are incomplete and can't be applied to all conditions, while the numerical tests \tianshu{in previous work} are not representative enough, since there are numerical examples \tianshu{we provide in this paper} \tianshu{for} which the monochromatic formulae will fail.}

In \citet{Lin2023}, the analytical growth rate of the minus mode in the multi-group case is derived under the assumption that ${\rm Re}(\Omega)\sim\mu D\equiv\sqrt{2}G_F(n_{\nu_e}-n_{\nu_x}-n_{\bar{\nu}_e}+n_{\bar{\nu}_x})\gg \Gamma(E)\sim {\rm Im}(\Omega)$, and this multi-group minus mode has the same form as that in the monochromatic case. However, the authors don't include the plus mode in multi-group cases, despite the fact that this assumption was not thoroughly tested. Hence, because of the missing plus modes, this derivation cannot be regarded as a complete justification for the use of the monochromatic formulae in multi-group cases. It is also worth noting that both $G$ and $\alpha$ are positive in CCSN environments, indicating that the missing plus mode is crucial.

\citet{Xiong2023} derives the analytical forms for both the plus and minus modes corresponding to the monochromatic formulae in multi-group cases. Similar to \citet{Lin2023}, \citet{Xiong2023} assumes\footnote{The actual assumption is $|\Omega|\gg|\omega_{\rm eff}|$ where $\omega_{\rm eff}=\cos(2\theta)\frac{\Delta m^2}{2E}+i\Gamma(E)$ and the vacuum term can be neglected.} $|\Omega|\gg\Gamma(E),\bar{\Gamma}(E)$ and expands the multi-group dispersion relation Eq. \ref{equ:dispersion_relation} to first order in $\frac{\Gamma(E)}{\Omega}$. The solutions are
\equ{
\Omega_\pm=\frac{\mu D}{2}\left(-1\pm\sqrt{1+\frac{4i\langle\Gamma\rangle}{\mu D}}\right)\, ,
}
where 
\equ{
\langle\Gamma\rangle=\frac{\int_{-\infty}^{\infty}E^2dE [f_{\nu_e}(E)-f_{\nu_x}(E)]\Gamma(E)}{\int_{-\infty}^{\infty}E^2dE [f_{\nu_e}(E)-f_{\nu_x}(E)]}\, .
\label{equ:xiong2023}
}
Here, we slightly modify the original definition of $\langle\Gamma\rangle$ in \citet{Xiong2023} to make it more consistent with our notation. When $\mu D\gg\langle\Gamma\rangle$, which is typically true in CCSNe, the solutions become $\Omega_+=i\langle\Gamma\rangle$ and $\Omega_-=-\mu D-i\langle\Gamma\rangle$. The authors find that the minus mode is exactly the same as that found in \citet{Lin2023} and the two solutions correspond to the two types of hCFI in the monochromatic neutrino gas. However, because $|\Omega_+|=\langle\Gamma\rangle\sim\Gamma(E)$, the plus mode in this limit typically violates the essential assumption $|\Omega|\gg\Gamma(E)$ used in the derivation. 
The only exception is the resonance case, in which $\int_{-\infty}^{\infty}E^2dE [f_{\nu_e}(E)-f_{\nu_x}(E)]$ is close to zero, so that $\langle\Gamma\rangle\propto\left(\int_{-\infty}^{\infty}E^2dE [f_{\nu_e}(E)-f_{\nu_x}(E)]\right)^{-1}$ diverges and becomes much larger than $\Gamma(E)$\footnote{Although $\mu D\propto \int_{-\infty}^{\infty}E^2dE [f_{\nu_e}(E)-f_{\nu_x}(E)]$, it is possible to find a case where $\mu D\gg\langle\Gamma\rangle\gg\Gamma(E)$ because $\mu D$ is typically many orders of magnitude higher than $\Gamma(E)$.}. Therefore, the justification for the plus mode solution is valid only in resonance-like hCFI. 

\citet{Liu2023} numerically studied the near-resonance behavior of the hCFI {and tests Method B}. However, the tested case was coincidentally in the valid context for the application of Method A in \citet{Xiong2023} {as discussed above}, {and it seems unsurprising that Method B will also behave nicely}. 
Therefore, the good performance of the monochromatic formulae in \citet{Liu2023} is \tianshu{not representative} and doesn't in itself justify the use of the monochromatic formulae in \tianshu{general} multi-group cases. 
In the Appendix, we show that the monochromatic formulae fail to give the correct growth rates if the prescribed parameters in \citet{Liu2023} are changed. Therefore, the reported good performance of Method B in \citet{Liu2023} lacks a general justification. 

Here, we give a more realistic example showing that the monochromatic formulae can mistakenly identify a strong instability when the neutrino gas is actually stable. The left panel of Figure \ref{fig:example} shows the normalized neutrino number spectra and the collision rates taken from our 18$M_\odot$ 1D reference simulation at a radius of 37 km and 120 milliseconds post-bounce. The local density, temperature, and $Y_e$ are $1.016\times10^{12}$ g cm$^{-3}$, 7.45 MeV, and 0.1426, respectively. Based on the monochromatic formulae, this location manifests the strongest hCFI growth rate at that time. The right panel shows the contour lines of ${\rm Re}({\rm I})=-1$ and ${\rm Im}({\rm I})=0$ on the ${\rm Re}(\Omega)$ and ${\rm Im}(\Omega)$ plane. The cross points of the contours give the solutions to the dispersion relation. The horizontal dashed line marks the hCFI growth rate calculated using the monochromatic formulae {(the results of Method A and B overlap on this scale so we only plot one line)}. It is clear from this example that all numerical solutions have negative imaginary parts and that there is no unstable mode. Although we show only the isotropy-preserving branch (I$=-1$), the same is true for the isotropy-breaking branch (I$=3$), because the growth rates are limited by the imaginary part of the equation (Im(I)=0). However, the monochromatic formula gives a growth rate of a few $10^4$ s$^{-1}$, which is qualitatively different from the correct answer. We note that this point is in the center of the hCFI region identified by the monochromatic formulae, and can't be consider an outlier. 
This example highlights a limitation of applying the monochromatic formulae to multi-group cases. Therefore, we suggest that all future hCFI studies in multi-group cases should numerically solve the dispersion relation Eq. \ref{equ:dispersion_relation}, unless they focus only on resonance-like hCFI.  \tianshu{In addition to the example we provide, \citet{Fiorillo+2025} recently has also shown in their Section 8.5 that the monochromatic formula could cause large deviations from the correct hCFI growth rates with certain parameterized neutrino spectra.}

\begin{figure*}[t]
    \centering
    \includegraphics[width=0.48\textwidth]{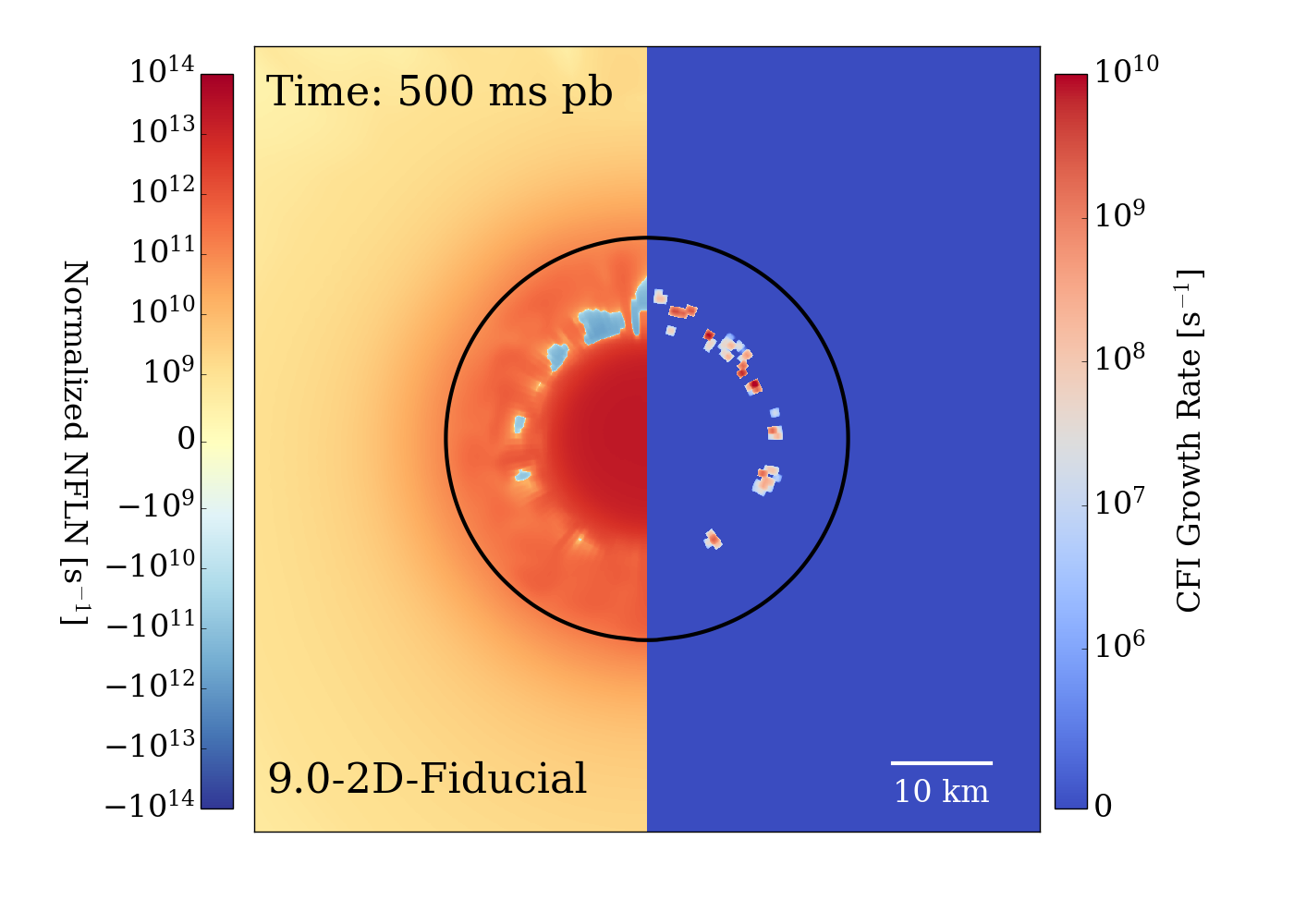}
    \includegraphics[width=0.48\textwidth]{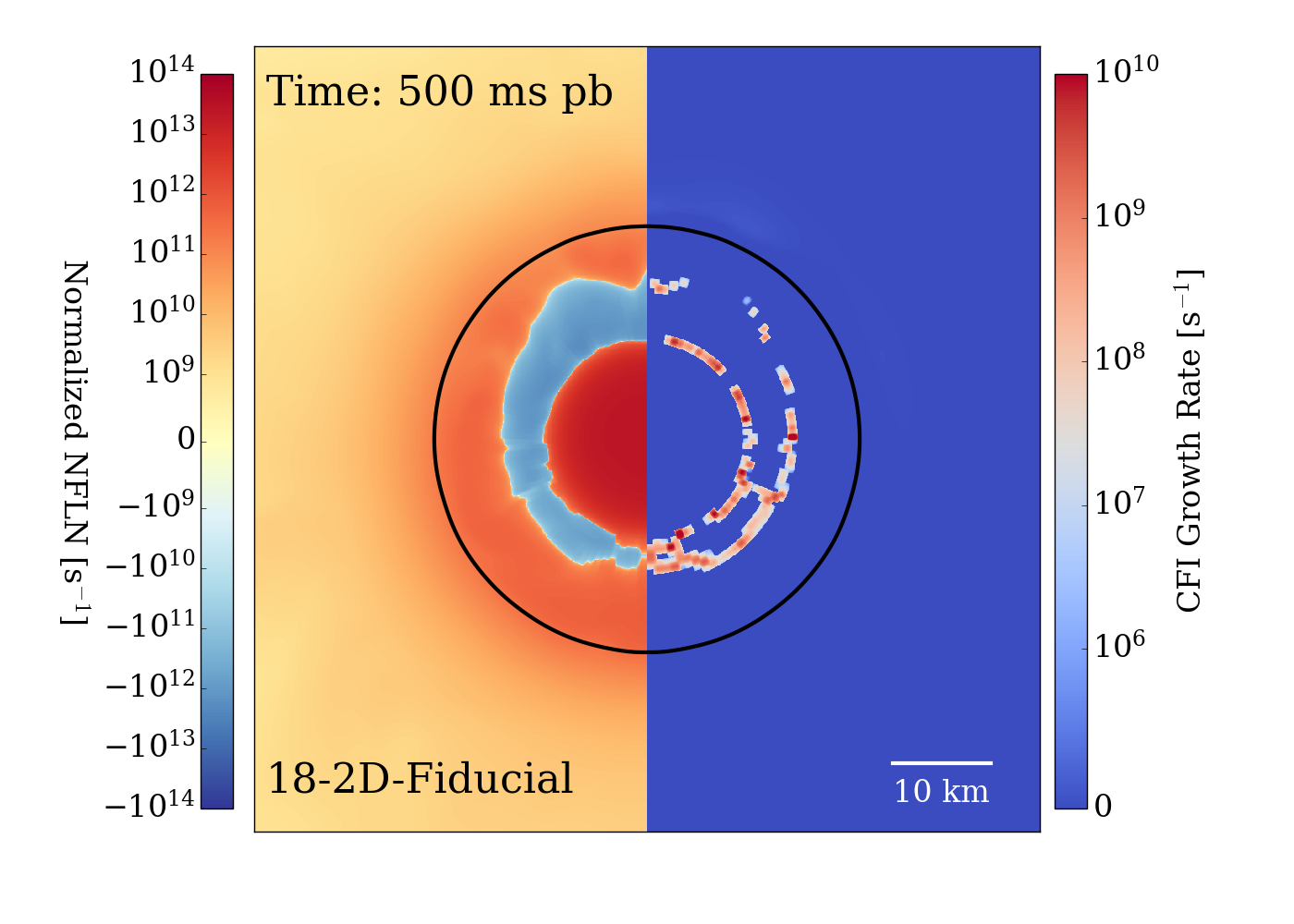}
    \caption{Normalized NFLN ($\sqrt{2}G_F(n_e-n_x-\bar{n}_e+\bar{n}_x)$) and hCFI growth rate snapshots of the 9 and 18 $M_\odot$ 2D fiducial models at 500 milliseconds post-bounce. The black contour shows the isosurface of $\rho=10^{13}$ g cm$^{-3}$. The left half of each panel shows the normalized NFLN, while the right half is the corresponding hCFI growth rate. The resonance condition requires NFLN to be equal to zero, which are in the thin layers located between the positive (red) and negative (blue) NFLN regions. The resonance-like hCFI occurs deep in the PNS convection region, interior to the $\rho=10^{13}$ g cm$^{-3}$ isosurface. We see larger resonance-like hCFI regions in the 18 $M_\odot$ model compared to the 9 $M_\odot$ model, which might be related to the weaker convection in the latter. The resonance-like hCFI growth rates are typically $10^{9}$ s $^{-1}$. Due to the finite spatial resolution in our simulations, some resonance-like hCFI region might be missed. However, as we will show later, resonance-like hCFI only weakly influences CCSN outcomes and the uncertainty due to the finite resolution is weak.}
    \label{fig:growth_rates_2d}
\end{figure*}

\section{Results}
\label{sec:result}
\subsection{Non-resonance hCFI}
As discussed above, the monochromatic formulae fail to give the correct plus mode growth rates in non-resonance cases when applied to multi-group systems. Since many previous works used the monochromatic formulae in multi-group cases \citep{Liu2023b,akaho2024}, it is important to revisit the identification of non-resonance hCFI regions in a representative CCSN context.

Figure \ref{fig:growth_rates} shows several hCFI growth rate profile snapshots of the 9, 12.25, 14, 18, 20, and 25 $M_\odot$ 1D fiducial models in the first 200 milliseconds post-bounce. Results calculated by the monochromatic formulae and by numerically solving the multi-group dispersion relation are plotted together for a clear comparison. In all snapshots shown here, the hCFI unstable regions are non-resonant. We find that the monochromatic formulae overestimate the maximum non-resonance hCFI growth rates by almost three orders of magnitudes {in all our models}. With the more accurate multi-group method, none of the models shows non-resonance hCFI growth rates above $\sim$200 s$^{-1}$. Since the size of these hCFI regions is at most a few tens of kilometers, the neutrino escape timescale through the unstable region ($\sim$10$^{-4}$ s) is much shorter than the timescale of hCFI growth\footnote{\tianshu{The identified non-resonance hCFI regions are mostly located above the neutrinosphere, thus $\Delta r/c$ provides a good estimation on the neutrino escape timescale.}}. As a result, the effect of flavor conversion triggered by such non-resonance hCFI is much weaker than the uncertainties introduced by other physical processes and can be ignored in current simulations.

\subsection{Resonance-Like hCFI}
Although the non-resonance hCFI grows too slowly, the higher growth rates of the resonance-like hCFI may be able to introduce noticeable effects in the properties of CCSN explosions. In our 1D fiducial models, we don't see any resonance-like hCFI up to 800 ms post-bounce. This absence is due to the artificial suppression of PNS convection under spherical symmetry, which limits the deleptonization of the PNS envelope \citep{Nagakura2020}. As a result, the chemical potential of $\nu_e$ remains higher than that of $\bar{\nu}_e$ thereby suppressing the occurrence of resonance-like hCFI. This is consistent with the findings in \citet{Liu2023b}, which also studied 1D models. This finding is expected because the monochromatic approximation they employ remains valid for the resonance-like hCFI, as discussed in Section \ref{sec:criticism}.

In 2D models, \citet{akaho2024} find that the resonance-like hCFI occurs in thin layers at matter densities higher than $\sim10^{13}$ g cm$^{-3}$. We confirm this finding and see resonance-like hCFI regions in all our 2D fiducial models. Figure \ref{fig:growth_rates_2d} shows the normalized NFLN (defined as $\sqrt{2}G_F\int_{-\infty}^{\infty}\frac{E^2dE}{2\pi^2} [f_{\nu_e}(E)-f_{\nu_x}(E)]/2=\sqrt{2}G_F(n_{\nu_e}-n_{\nu_x}-n_{\bar{\nu}_e}+n_{\bar{\nu}_x})/2$) and the corresponding resonance-like hCFI growth rate calculated by the monochromatic formulae. Note that the monochromatic formulae are able to give the correct answer in the resonance-like case as discussed in Section \ref{sec:criticism} and \citet{Liu2023}. The resonance-like hCFI regions are very thin layers located in the inner PNS convection region, and the local matter density is always above $10^{13}$ g cm$^{-3}$ throughout our simulations. We see larger resonance-like hCFI regions in the 18 $M_\odot$ model compared to the 9 $M_\odot$ model, which might be related to the weaker convection in the latter. With the finite spatial resolution in F{\sc ornax} simulations, the resonance-like hCFI growth rates are about $10^{9}$ s$^{-1}$. Although the growth rates don't go to infinity, they are high enough for the flavor asymptotic states to be achieved in one single timestep (again, generally equal to $\sim10^{-6}$ s). 

\begin{figure*}[t]
    \centering
    \includegraphics[width=0.48\textwidth]{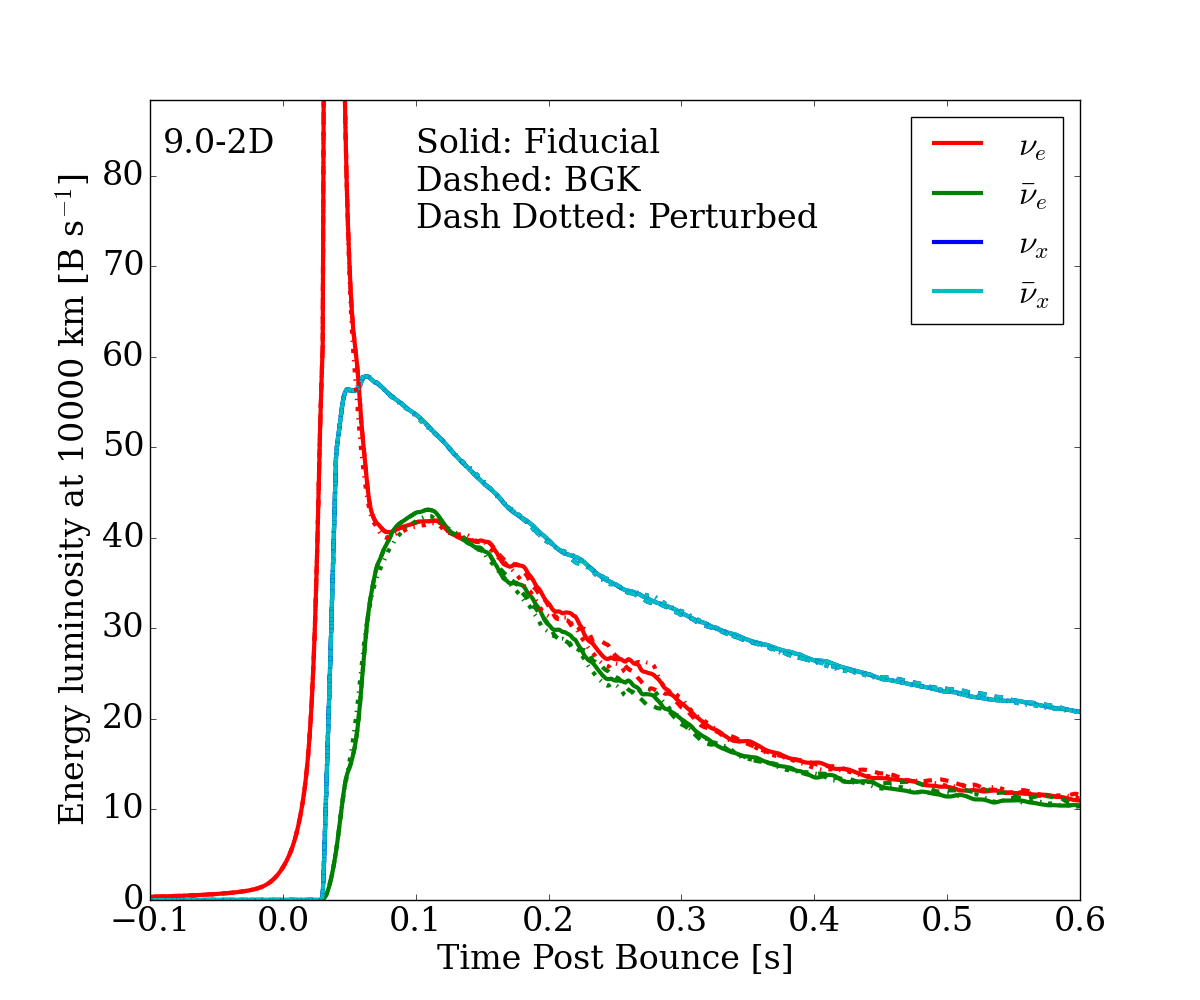}
    \includegraphics[width=0.48\textwidth]{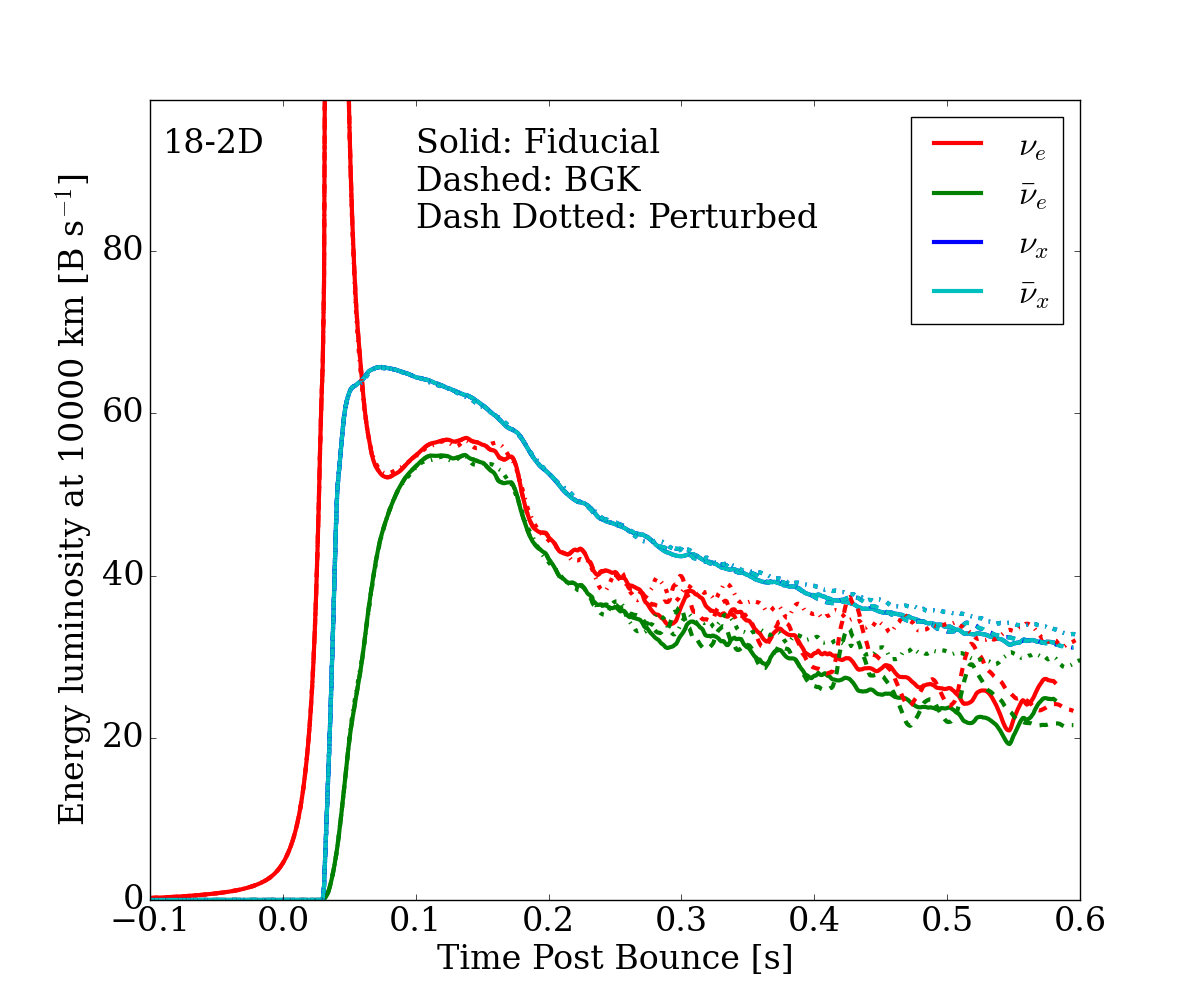}
    \includegraphics[width=0.48\textwidth]{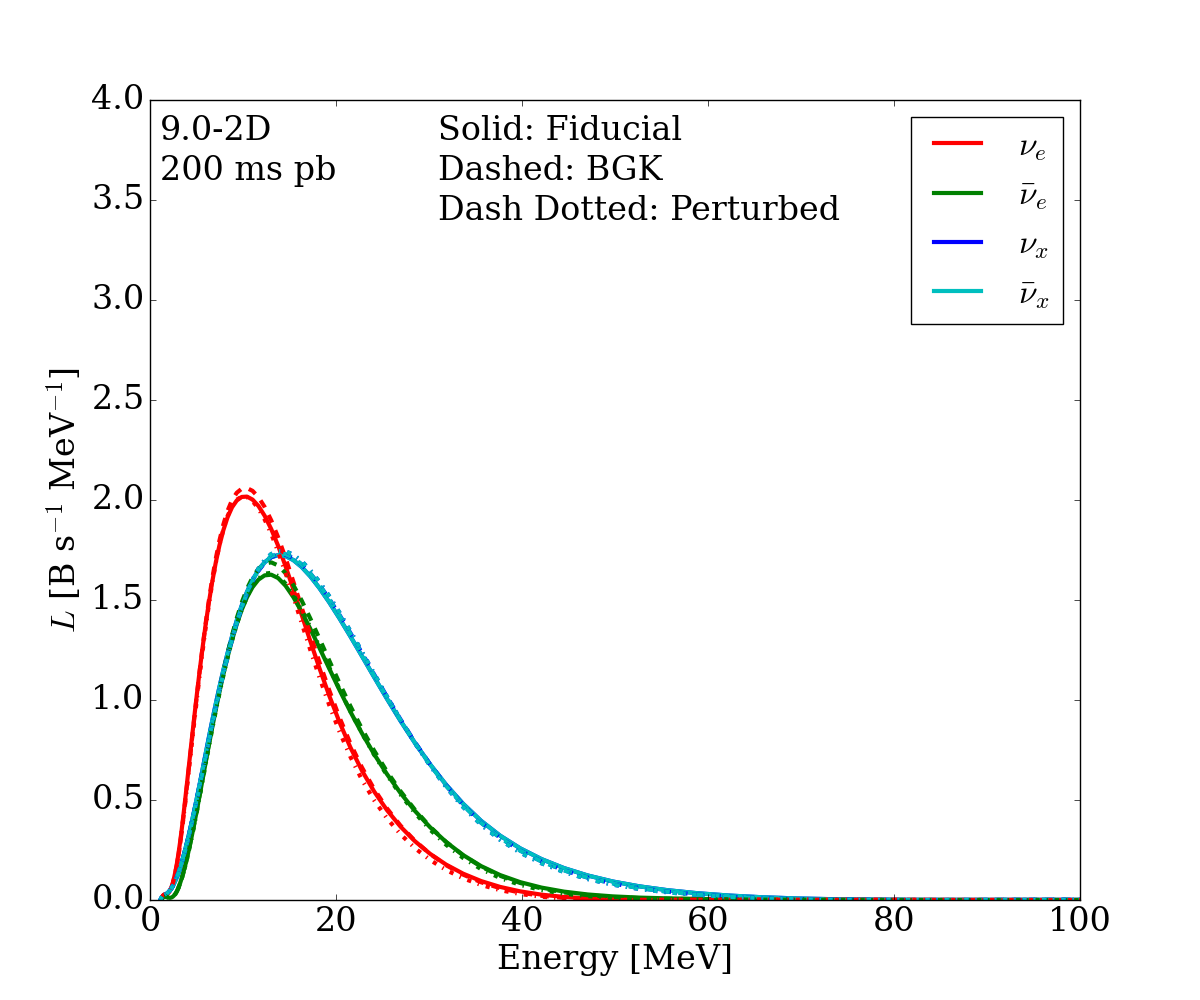}
    \includegraphics[width=0.48\textwidth]{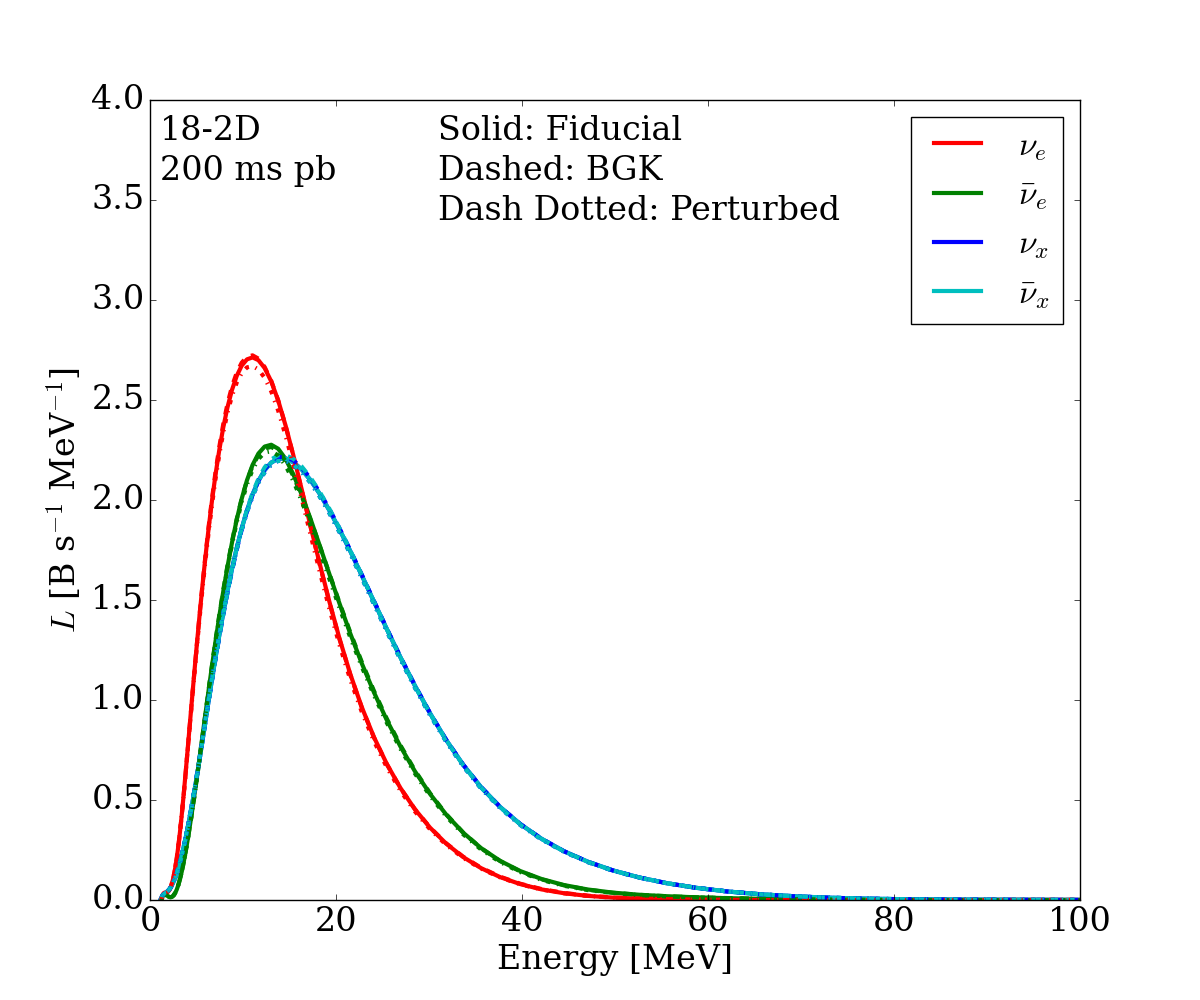}  
    \caption{Neutrino emission properties measured at 10000 km of the 9 and 18 $M_\odot$ models. \tianshu{To show the intrinsic stochastic effects of CCSNe, we also include ``perturbed'' models, which are fiducial models plus $10^{-6}$ random density perturbations at the beginning of the simulations.} {\bf Top:} Neutrino luminosities as a function of time. {\bf Bottom:} Neutrino spectra at 200 ms post-bounce. Only resonance-like hCFI is considered in these BGK models because the non-resonance hCFI can be ignored due to its low growth rate. Differences between models with and without the hCFI effects are minor. At relatively late times, variations in the electron and anti-electron type neutrino luminosities become larger. However, we think this is not directly related to flavor conversion, but due to the natural chaos in the solution in turbulent flow. If the slight variations seen for the $\nu_e$ and $\bar{\nu}_e$ neutrinos were due to flavor conversion, there should be a similar degree of variation in the $\nu_x$ and $\bar{\nu}_x$ neutrinos, which is not seen. Therefore, as suggested, the variations are more likely due to the intrinsic stochasticity of CCSN explosions.  The hCFI effects are much weaker \tianshu{than the effects caused by a weak initial random perturbation field}.}
    \label{fig:neutrino}
\end{figure*}

We select the 9 and 18 $M_\odot$ models as two representatives and redo the simulations with the BGK flavor conversion scheme. Figure \ref{fig:neutrino} shows the neutrino emission properties measured at 10000 km. \tianshu{To show the intrinsic stochastic effects of CCSNe, we also include ``perturbed'' models, which are fiducial models plus $10^{-6}$ random density perturbations at the beginning of the simulations.} Only minor differences are seen between models with and without the hCFI effects. At relatively late times, variations in the electron and anti-electron type neutrino luminosities become larger. However, we think this is not directly related to flavor conversion. If the variations in $\nu_e$ and $\bar{\nu}_e$ are due to flavor conversion, there should be similar amounts of variation for $\nu_x$ and $\bar{\nu}_x$ neutrinos, which is not seen. Therefore, the variations are more likely due to the intrinsic stochasticity (convection, turbulence, explosion asymmetry, etc.) of CCSN explosions and the effects of hCFI are significantly weaker \tianshu{than the effects caused by a weak initial random perturbation field}. 

\tianshu{Since the BGK scheme instantly reaches flavor equipartition when growth rates are sufficiently high, the outcomes of these models don't depend on the actual type of flavor instability. Therefore, our models show that the effects of any flavor instability triggered only by vanishing NFLN are minor. One example of such instabilities is the fast flavor instability found at vanishing NFLN locations in \citep{Glas2020}.}

\section{Conclusion}
\label{sec:conclusion}
In this paper, we presented a comprehensive analysis of the flavor conversion effects triggered by hCFI modes in 1D and 2D CCSN simulations between 9 and 25 $M_\odot$. First, we pointed out that the widely-used monochromatic formulae for hCFI growth rates are problematic in the realistic multi-group context and that the multi-group dispersion relation should be solved numerically instead. The maximum non-resonance hCFI growth rates are systematically overestimated by the monochromatic formulae by nearly three orders of magnitudes in all our CCSN simulations. Thus, the previously identified non-resonance hCFI regions in \citet{Liu2023b,akaho2024} and the inferred effects on CCSN dynamics and the neutrino fields would be in error. 

The resonance-like hCFI, on the other hand, is still well-approximated by the monochromatic formulae. We confirm the findings in \citet{Liu2023b,akaho2024} that resonance-like hCFI occurs only in multi-dimensional simulations at relatively late times. The properties of the resonance-like hCFI regions are consistent with those found in \citet{akaho2024}. Although the physical sizes of resonance-like hCFI unstable regions are very small, the diverging growth rates preserve the possibility to significantly mix different neutrino flavors.

To understand the resonance-like hCFI effects on CCSN dynamics, we adopted a BGK flavor conversion scheme in the supernova code F{\sc ornax} and redid the 9 and 18 $M_\odot$ 2D models. We found that the shock radii and the neutrino emission properties are changed by at most a few percent. This is a weak effect, because the intrinsic stochasticity due to convection and neutrino-driven turbulence can naturally lead to comparable effects. The remaining uncertainties in the nuclear equation of state \citep{couch2013,yasin2020,Boccioli2022} and in the structures of the initial model progenitors \citep{swbj16,sukhbold2018} likely swamp this effect.    
Hence, our analysis of the non-resonance and resonance-like hCFI into CCSN simulations suggests that the effects of neutrino flavor conversion triggered by hCFI modes are in general small.

We wrap up this paper by summarizing the limitations to this work. First, our analysis is based on a linear analysis of hCFI unstable modes in a homogeneous and isotropic neutrino background. Although this assumption is true in very optically thick regions, the non-resonance hCFI can occur in semi-transparent regions where neutrino angular distributions deviate from isotropic ones. It is unclear whether an anisotropic and/or inhomogeneous neutrino background will change our conclusions. The effects of inhomogeneous {CFI} modes with non-vanishing wave vectors remain to be studied. Second, the effects of flavor conversion are handled by a BGK method with prescribed asymptotic states. We assume that the asymptotic state of flavor conversion is flavor equipartition, but the actual outcome might be more complex \citep{kato2024,Zaizen2025,Froustey2025}. Although our conclusion seem not to be sensitive to the choice of asymptotic states due to the overall weak effects of hCFI, it will be useful to improve such subgrid models in future. In addition, we use a 4-species scheme in this work, while a full 6-species scheme is desirable. Also, we have ignored the possible presence of muons, and we assume that $\nu_x$ and $\bar{\nu}_x$ neutrinos have the same emissivities/opacities, which suppresses CFI in the $\mu$-$\tau$ sector \citep{Liu2024}. Moreover, the effects of fast and slow flavor instabilities are not included in this work. It is unclear if different types of instabilities may interact with each other and change our conclusions. Furthermore, our analysis is based on F{\sc ornax} simulations and the microphysics included. We notice that the multi-group analysis of a 18 $M_\odot$ 1D model in \citet{Xiong2023b} leads to higher {CFI} growth rates. Potential reasons for the discrepancy include inhomogeneous modes, anisotropy of background neutrinos, different CCSN models, etc. Further studies are called for.

\section*{Acknowledgment}
We thank David Vartanyan for our long-term productive collaboration and Daniel Kasen for the many discussions on the {CFI}-related CCSN simulations. TW acknowledges support by the U.~S.\ Department of Energy under grant DE-SC0004658, support from the Gordon and Betty Moore Foundation through Grant GBMF5076 and through Simons Foundation grant (622817DK). H.N. is supported by Grant-in-Aid for Scientific Research (23K03468),
the NINS International Research Exchange Support Program, and the HPCI
System Research Project (Project ID: hp250006, hp250226, hp250166). LJ is supported by a Feynman Fellowship through LANL LDRD project number 20230788PRD1. AB acknowledges former support from the U.~S.\ Department of Energy Office of Science and the Office of Advanced Scientific Computing Research via the Scientific Discovery through Advanced Computing (SciDAC4) program and Grant DE-SC0018297 (subaward 00009650) and former support from the U.~S.\ National Science Foundation (NSF) under Grant AST-1714267. We are happy to acknowledge access to the Frontera cluster (under awards AST20020 and AST21003). This research is part of the Frontera computing project at the Texas Advanced Computing Center \citep{Stanzione2020}. Frontera is made possible by NSF award OAC-1818253. Additionally, a generous award of computer time was provided by the INCITE program, enabling this research to use resources of the Argonne Leadership Computing Facility, a DOE Office of Science User Facility supported under Contract DE-AC02-06CH11357. Finally, the authors acknowledge computational resources provided by the high-performance computer center at Princeton University, which is jointly supported by the Princeton Institute for Computational Science and Engineering (PICSciE) and the Princeton University Office of Information Technology, and our continuing allocation at the National Energy Research Scientific Computing Center (NERSC), which is supported by the Office of Science of the U.~S.\ Department of Energy under contract DE-AC03-76SF00098.

\appendix*
\section{}
\label{sec:appendix}

Here, we first reproduce the numerical results of \citet{Liu2023} and then provide a few examples in which the monochromatic formulae fail to give the correct growth rates.

The problem setup in \citet{Liu2023} is:
\equ{
&&f_{\nu_e}=\left[\exp\left(\frac{E}{4{\rm MeV}}\right)+1\right]^{-1}\nonumber\\
&&f_{\bar{\nu}_e}=\bar{g}\left[\exp\left(\frac{E}{5{\rm MeV}}\right)+1\right]^{-1}\nonumber\\
&&f_{\nu_x}=f_{\bar{\nu}_x}=0\nonumber\\
&&\Gamma=\Gamma_0\left(\frac{E}{10{\rm MeV}}\right)^2\nonumber\\
&&\bar{\Gamma}=\bar{\Gamma}_0\left(\frac{E}{10{\rm MeV}}\right)^2\, ,
}
where $\bar{g}$, $\Gamma_0$, and $\bar{\Gamma}_0$ are parameters. In \citet{Liu2023}, $\Gamma_0$ and $\bar{\Gamma}_0$ are both fixed to $3\times10^5$ s$^{-1}$ and the only parameter which changes is $\bar{g}$
\footnote{We multiply the original $\Gamma$ and $\bar{\Gamma}$ in \citet{Lin2023} by the speed of light so that the unit of the rates is s$^{-1}$ rather than cm$^{-1}$, which is consistent with our notation. This unit conversion doesn't influence the growth rate calculation.}. In our tests, the energy range from 0.01 MeV to 100 MeV is divided uniformly into 200 bins, which is a higher resolution than \citet{Liu2023}.

\begin{figure*}
    \centering
    \includegraphics[width=0.48\textwidth]{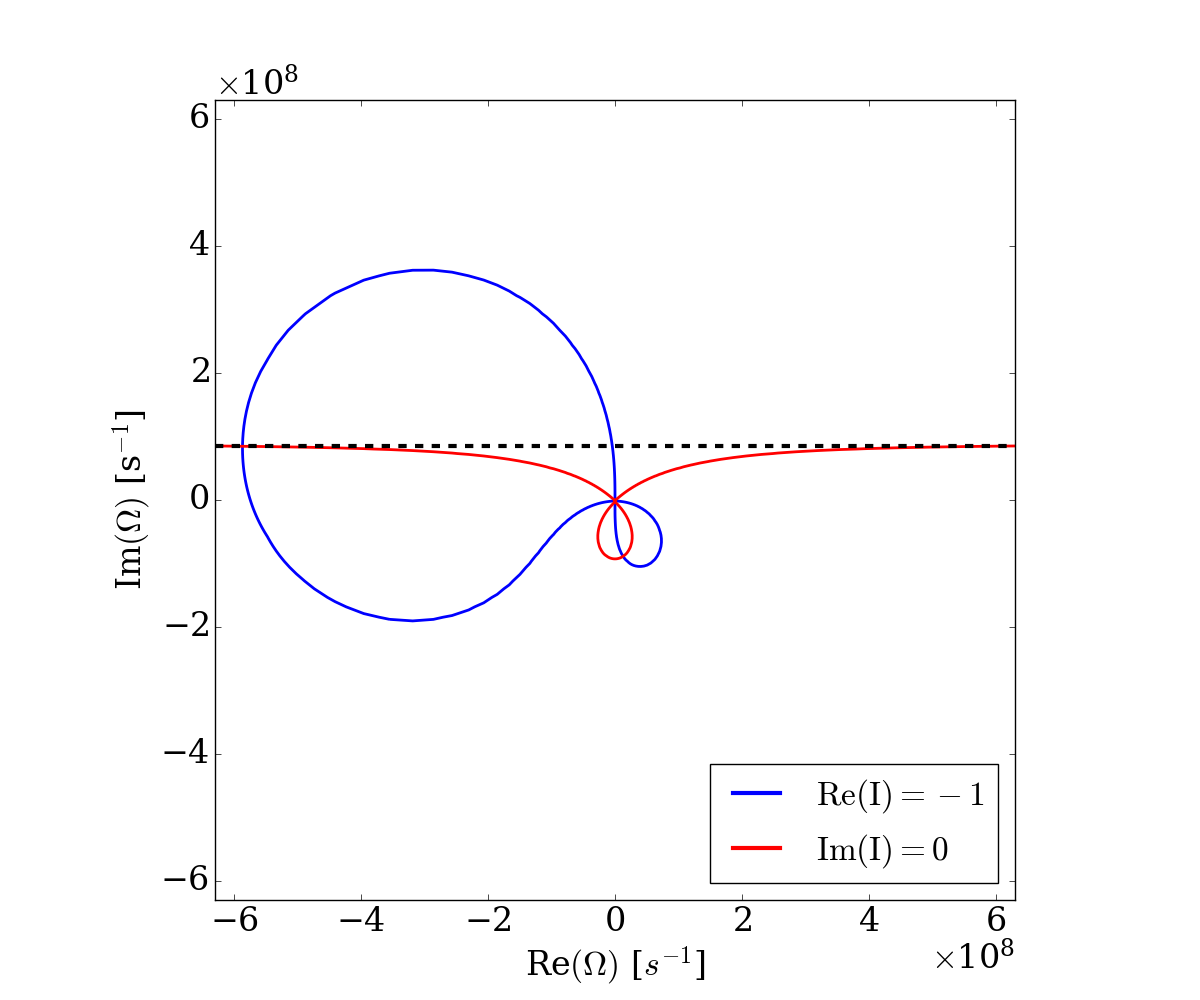}
    \includegraphics[width=0.48\textwidth]{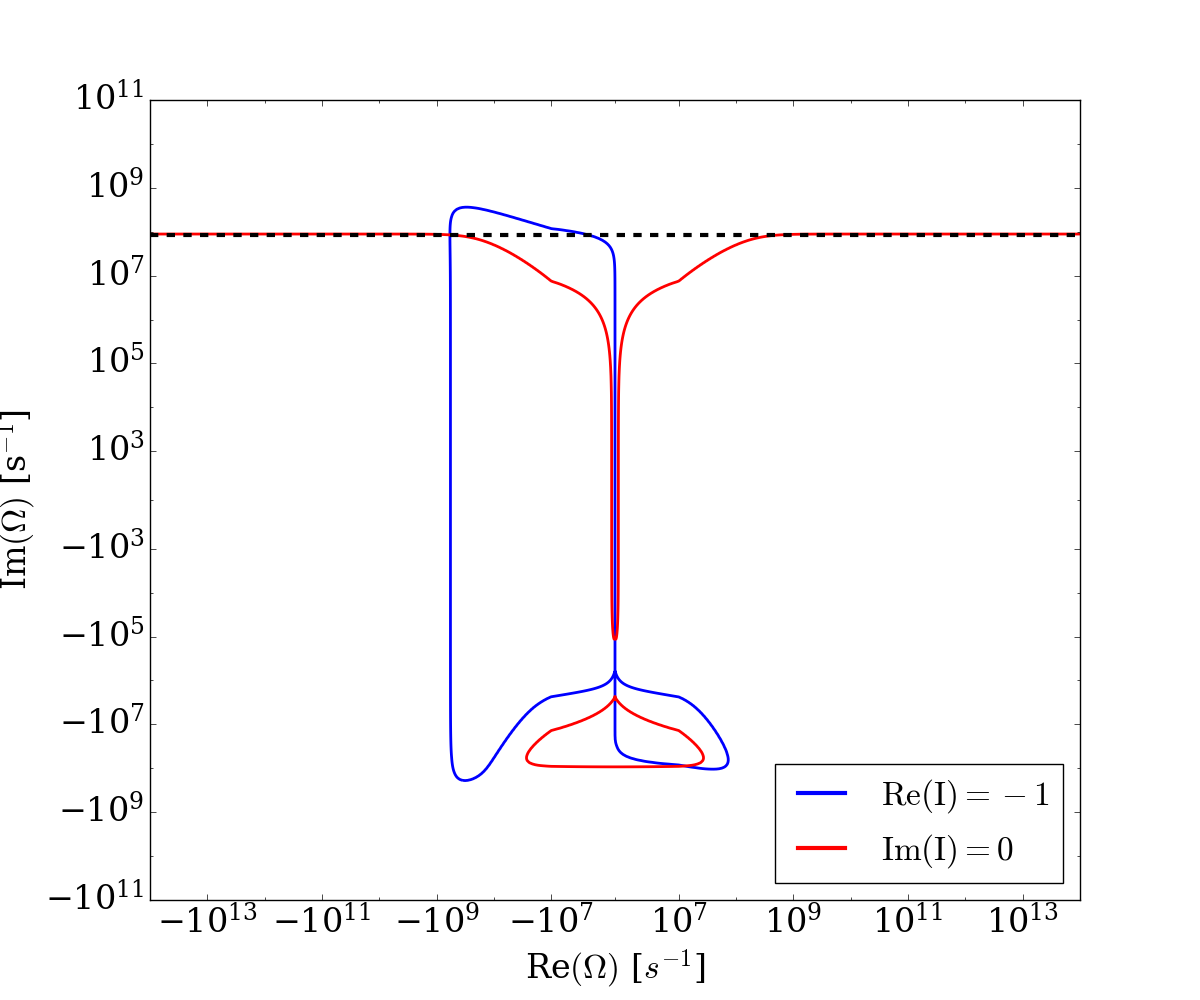}
    \caption{The reproduction of the $\bar{g}=0.51$ case in \citet{Liu2023}. The red and blue contour lines mark Re(I)=$-$1 and Im(I)=0 in Eq. \ref{equ:dispersion_relation}. The cross points of the contours give the solutions to the dispersion relation. The horizontal dashed line depicts the maximum monochromatic growth rate. The left and right panels are identical, except that the right panel uses the symlog scale. The horizontal dashed line shows the monochromatic growth rate calculated by Method B, which is the method tested in \citet{Liu2023}. The left panel is identical to Figure 5(a) of \citet{Liu2023}, showing that our numerical method is consistent with those used in previous work.}
    \label{fig:reproduction}
\end{figure*}

Figure \ref{fig:reproduction} shows the reproduction of $\bar{g}=0.51$ case in \citet{Liu2023}. The left and right panels are identical except that the right panel uses the symlog scale. The horizontal dashed line shows the monochromatic growth rate calculated by Method B. The left panel is identical to Figure 5(a) of \citet{Liu2023}, showing that our numerical method is consistent with those used in previous work. We have done similar tests for $\bar{g}=0.511$ and 0.512, and the results are identical to the corresponding cases in \citet{Liu2023}.

\begin{figure*}
    \centering
    \includegraphics[width=0.48\textwidth]{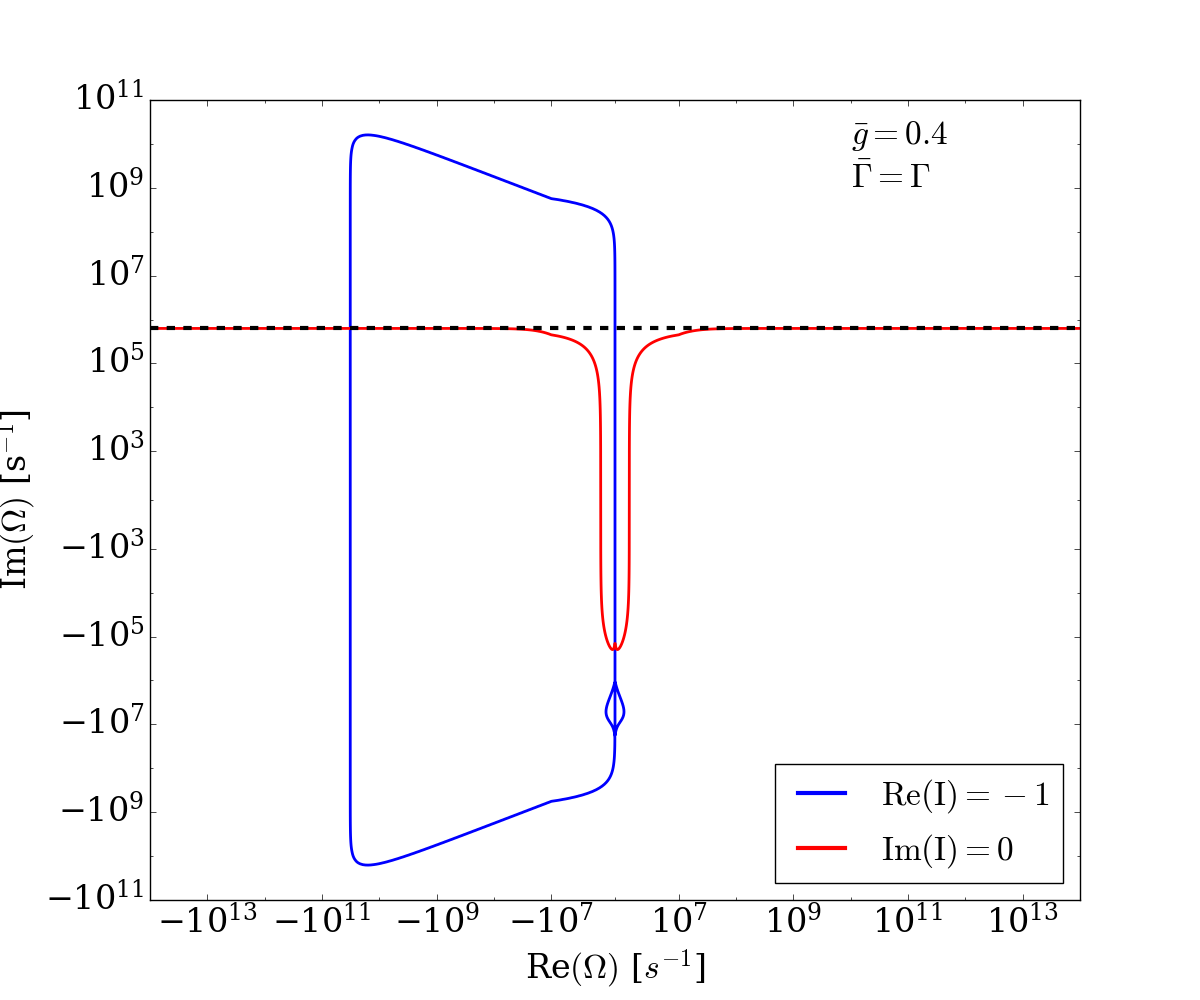}
    \includegraphics[width=0.48\textwidth]{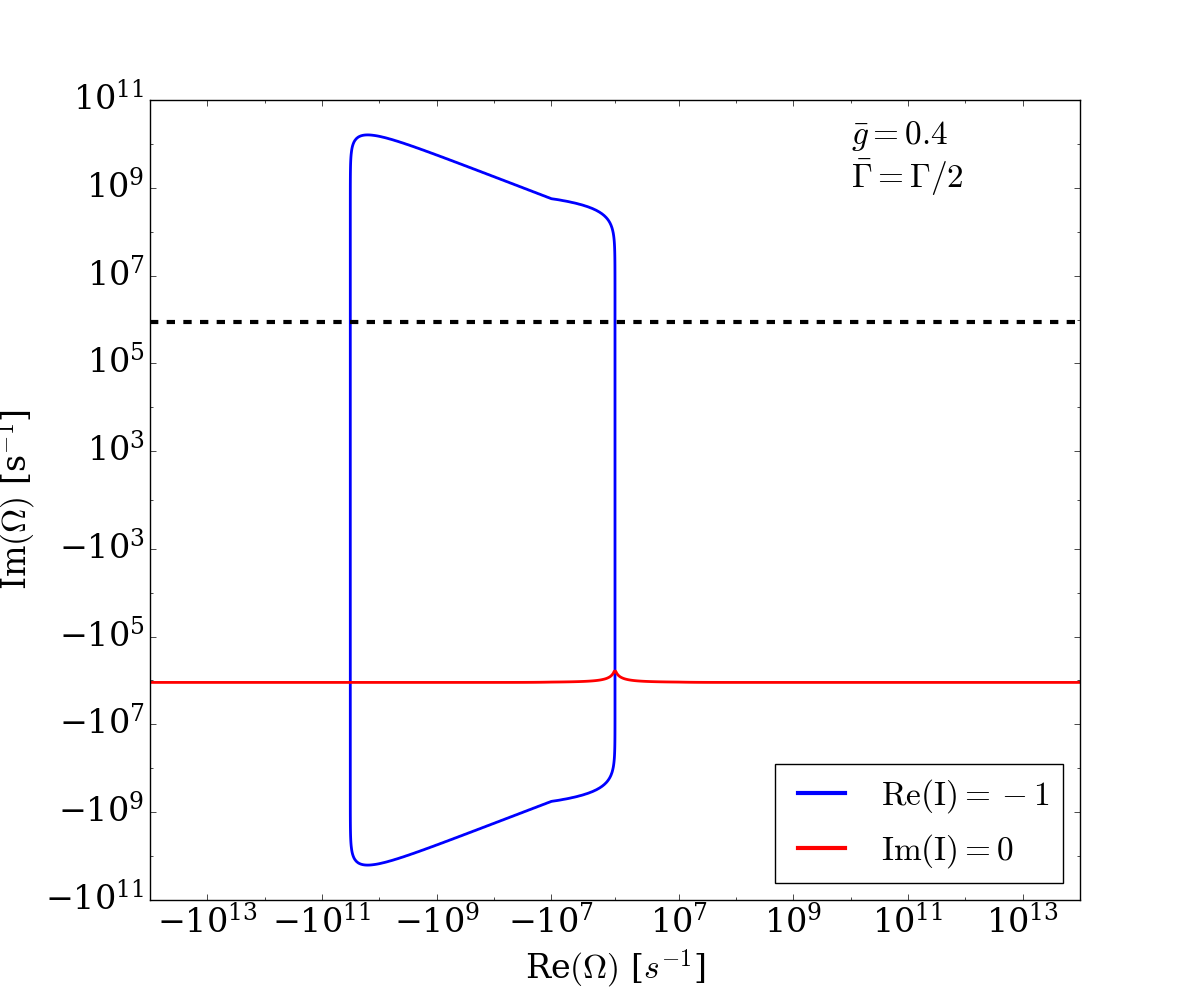}
    \caption{Same as Figure \ref{fig:reproduction}, but for non-resonance test cases. Here we set $\bar{g}=0.4$ and test various $\bar{\Gamma}_0$. We find that when $\Gamma_0$ and $\bar{\Gamma_0}$ are equal to each other, as done in \citet{Liu2023}, the monochromatic formulae give the correct growth rates. However, if we set $\Gamma$ and $\bar{\Gamma}$ to different values, the monochromatic results start to deviate from the correct solution. 
    }
    \label{fig:failure}
\end{figure*}

Figure \ref{fig:failure} shows the non-resonance test cases. We set $\bar{g}$ to 0.4 so that it's sufficiently far from the resonance ($\bar{g}=0.512$). Then we test two different $\bar{\Gamma}_0$'s:  $\bar{\Gamma}_0=\Gamma_0$ and $\bar{\Gamma}_0=\Gamma_0/2$
We find that the monochromatic formulae give the correct result in the $\bar{\Gamma}_0=\Gamma_0$ case, but in the $\bar{\Gamma}_0=\Gamma_0/2$ case the monochromatic rate is incorrect.
Since the collision rates $\Gamma_0$ and $\bar{\Gamma}_0$ are typically different in realistic environments, we suggest that monochromatic formulae can't be used in multi-group analysis.

\bibliography{CFI}

\clearpage

\end{document}